\documentclass[usenatbib]{mn2e}
\input{epsf}
\usepackage{amssymb}
\usepackage{graphicx}
\usepackage{natbib}
\usepackage{color}
\usepackage{journals}

\newbox\grsign \setbox\grsign=\hbox{$>$} \newdimen\grdimen \grdimen=\ht\grsign
\newbox\simlessbox \newbox\simgreatbox
\setbox\simgreatbox=\hbox{\raise.5ex\hbox{$>$}\llap
     {\lower.5ex\hbox{$\sim$}}}\ht1=\grdimen\dp1=0pt
\setbox\simlessbox=\hbox{\raise.5ex\hbox{$<$}\llap 
     {\lower.5ex\hbox{$\sim$}}}\ht2=\grdimen\dp2=0pt

\newcommand{\hMpc}{{\ifmmode{h^{-1}{\rm Mpc}}\else{$h^{-1}$Mpc }\fi}}
\newcommand{\hkpc}{{\ifmmode{h^{-1}{\rm kpc}}\else{$h^{-1}$kpc }\fi}}
\newcommand{\hMsun}{{\ifmmode{h^{-1}{\rm {M_{\odot}}}}\else{$h^{-1}{\rm{M_{\odot}}}$}\fi}}
\newcommand{\Msun}{{\ifmmode{{\rm {M_{\odot}}}}\else{${\rm{M_{\odot}}}$}\fi}}

\title[Voids over cosmic time]{Voids in cosmological simulations over cosmic time} 
\author[R. Wojtak]{Rados{\l}aw Wojtak$^{1,2}$\thanks{E-mail: wojtak@stanford.edu}, Devon
  Powell $^{1}$, Tom Abel$^{1}$
  \\   \\
  $^1$Kavli Institute for Particle Astrophysics and Cosmology, Stanford University, 
  SLAC National Accelerator Laboratory, \\
  Menlo Park, CA 94025, USA\\
  $^2$Dark Cosmology Centre, Niels Bohr Institute, University of
  Copenhagen, Juliane Maries Vej 30, \\ DK-2100 Copenhagen \O,
  Denmark\\
}

\begin{document}

\maketitle

\begin{abstract}

We study evolution of voids in cosmological simulations using a new method for tracing 
voids over cosmic time. The method is based on tracking watershed basins 
(contiguous regions around density minima) of well developed 
voids at low redshift, on a regular grid of density field. It enables us to construct a robust and continuous mapping 
between voids at different redshifts, from initial conditions to the present time. We discuss how the new approach 
eliminates strong spurious effects of numerical origin when voids evolution is traced by matching voids between 
successive snapshots (by analogy to halo merger trees). We apply the new method to a cosmological 
simulation of a standard $\Lambda$CDM cosmological model and study evolution of basic properties of typical voids 
(with effective radii $6\hMpc<R_{\rm v}<20\hMpc$ at redshift $z=0$) such as volumes, shapes, matter density distributions 
and relative alignments. The final voids at 
low redshifts appear to retain a significant part of the configuration acquired in initial conditions. Shapes of voids 
evolve in a collective way which barely modifies the overall distribution of the axial ratios. The evolution appears 
to have a weak impact on mutual alignments of voids implying that the present state is in large part set up 
by the primordial density field. We present evolution of dark matter density profiles computed on isodensity surfaces 
which comply with the actual shapes of voids. Unlike spherical density profiles, this approach enables us to demonstrate 
development of theoretically predicted bucket-like shape of the final density profiles indicating a wide flat core 
and a sharp transition to high-density void walls.

\end{abstract}


\section{Introduction}

Cosmic voids are the most widespread and faintest structures in the Universe. They occupy around $80$ percent of space 
containing only $15$ per cent of the total mass \citep{Fal2015,Cau2014}. Driven by current and upcoming massive cosmological surveys, 
a growing interest in voids enters a phase of studies aimed at exploring various prospects of using voids as a complementary means for 
probing the nature of dark energy \citep{Lav2012}, dark matter \citep{Hel2009,Yan2015} or gravity \citep{Li2012,Cai2015} 
on large scales. Despite this promising potential of voids as cosmological probes, we lack a comprehensive 
understanding of their evolution and the origin of their properties. 


The basic framework for understanding evolutionary processes of cosmic voids was introduced by \citet{She2004} on the basis 
of excursion set theory \citep[see also the review of][]{Wey2011}. Unlike dark matter haloes, voids can either grow over time 
or can be destroyed through the gravitational collapse of the boundary walls. Which evolutionary paths voids follow depends on 
void sizes and their position in the void hierarchy. Small distinct voids tend to collapse, whereas large voids undergo expansion. 
The expansion of voids render their interior similar to finite open universes with self-similar structures of their cosmic web 
\citep{Got2003,Ara2010}. The evolution of the hierarchical void network was studied in numerical simulations \citep{Wey1993}, 
in the context of the adhesion approximation \citep{Sah1994} and excursion set theory \citep{She2004}. Most approaches 
focus on evolution of various statistics of voids such as void abundances \citep{Sut2014c} or overall distributions of such 
properties as sizes or shapes \citep{Bos2012}. However, considering overall statistics may conceal many interesting 
aspects of evolution of individual voids. As we shall demonstrate in this work, tracing evolution of individual voids has 
potential to unveil more complex and richer reality of evolutionary processes than tracing evolution of an overall statistics 
of a void population as a whole.

Tracing the evolution of individual voids in cosmological simulations requires using a robust method for tracking these objects 
between snapshots. One can attempt to do this by applying standard algorithms for constructing 
merger trees of dark matter haloes. Voids are found at different moments of simulations by means of tracking IDs of 
particles found in their interiors. This approach was adopted by \citet{Sut2014c} who used it to study formation, evolution and 
destruction of voids in cosmological simulations. Despite its straightforward implementation, the method is not expected to 
be accurate enough to separate genuine void evolution from a number of artificial effects related to the way of generating 
void catalogues and using particles as void tracer. In this work, we analyse performance of this approach and propose 
a new method which circumvents a number of recognized problems. The new method of tracking voids in simulations 
is built upon some generic properties of voids found with methods employing the watershed transform \citep{Pla2007,Ney2008}.

The current-state cosmic web was already present in the initial fluctuations of the primordial density field \citep{Bon1996}. 
Therefore, it is natural to think that the present voids can be mapped into underdense regions around initial troughs 
of the primordial density field \citep{Wey1993}. Remarkable resemblance between initial and final cosmic web suggests that initial 
underdense regions should arrange most of the properties of well-developed voids such as shapes and relative alignments. 
On the other hand, one should also expect important effects from a late-time non-linear evolution driven by tidal forces and 
constraints resulting from evolution of adjacent voids \citep{Pla2008}. However, it is not clear to what degree these effects 
erase the initial arrangement of the void network and to what degree voids retain their initial properties. In our work, we address 
this nature-nurture problem for the first time in a fully quantitative way. Applying a newly developed method for tracking voids 
in cosmological simulations we study how non-linear evolution affects the properties of voids acquired from initial troughs 
in the primordial density field.

Voids are traditionally defined as parts of space with the mean density below a certain fixed limit. The most commonly adopted 
density limit is $20$ per cent of the background density. The limit is derived in an analytic model of void evolution and signifies a moment 
of the first shell crossing leading to the formation of void boundaries \citep{She2004}. Despite a deep theoretical motivation, 
this approach narrows substantially the definition of voids. In particular, it automatically imposes the void formation time as the moment 
when void's density minimum crosses the assumed density threshold \citep{Sut2014c}. However, the process 
of void formation should be rather regarded as continuous in the course of time. Recalling the fact that the overall arrangement 
of the cosmic web is already present 
in the primordial density field, we should think of void formation as a gradual process increasing a density contrast between 
void cores and boundary walls. Following this picture we define voids as distinct underdense regions regardless 
of their evolutionary stage and thus regardless of their minimum or mean density. This simple definition allows us also to 
avoid a semantic ambiguity of possible terms which would have to be introduced in order to distinguish 
voids after formation time from void progenitors or protovoids before that time.

The paper is organized as follows. In Section 2 we describe all elements of the void finder: density estimator, the watershed algorithm 
and the method of generating void catalogues. In Section 3 we describe the new method of tracking voids in cosmological simulations and 
compare it to results from applying a direct implementation of a halo merger tree generator. We apply the new method to a standard 
cosmological simulation in Section 4, where we quantify evolution of various properties of individual voids (sizes, shapes, alignments 
and dark matter density distribution) from initial conditions to the present time. We conclude and summarize in Section 5.

\section{Void finder}
Voids can be found in various ways \citep[see e.g.][]{Col2008}. Many methods define voids as regions devoid of dark matter \citep{Pla2007} 
or dark matter haloes \citep{Got2003}, other methods utilize kinematic or dynamical features such as divergence 
of velocity field \citep{Hof2012} or tidal instability in smoothed density field \citep{Hah2007,Cau2013}. 
Some methods impose certain properties of voids such as shapes or density threshold, other techniques 
avoid fixing any of them and find voids in a non-parametric way. Void finders shall be also distinguished 
with respect to whether they can be applied to observational data or not. In general, void finders which 
need more sophisticated information than halo or galaxy distribution have more limited applicability to observations.

In our study we find voids as underdense regions in dark matter density field. Without attempting to show 
superiority of this approach over other methods, we emphasize that void finders based on this 
operational definition take a leading role in studying voids both in cosmological simulations and observations. 
Two most commonly used void finders, ZOBOV \citep[ZOnes Bordering On Voidness;][]{Ney2008} 
and its alteration VIDE \citep[the Void Identification and Examination tool;][]{Sut2015}, 
are arguably the most successful implementations \citep[see also][]{Way2015}. Voids found as underdense regions appear to be 
visually appealing elements of the cosmic web. Some void finders such as ZOBOV can be run in a 
non-parametric way that is essential to avoid imposing some preconceptions on what voids should look like. 
The same void finders are also capable of unveiling hierarchical network of voids which is 
one of the most characteristic properties of these objects \citep[see e.g.][]{Ara2010,Ara2013}.

We build our void finder by full analogy to ZOBOV. The only difference between our implementation and 
the original code is the way we compute the density field. Instead of a Voronoi diagram employed in ZOBOV, 
we use a Lagrangian Tessellation Field Estimator on a regular grid \citep{Abel2012}. 
The two remaining parts of the void finder, i.e. the watershed transform \citep{Pla2007} and 
a method of building physical voids, remain unchanged. We use regular density grid in order to 
optimize finding evolution of voids over successive snapshots of a simulation. 
Bearing in mind that voids retain their initial volumes rather than masses, 
it is natural to define voids as sets of pixels in space rather than a set of particles. 
In the latter case (for Voronoi tessellation), identification of voids in different snapshots becomes more ambiguous due 
to the fact that particles occupy void boundaries (walls) and therefore they are likely exchanged 
between adjacent voids. In the following subsections we describe all parts of our void finder in more detail.

\subsection{Density estimator}

We calculate the density field using the novel approach proposed by \citet{Abel2012} and
\citet{Sha2012} alongside the exact conservative voxelization method of \citet{Pow2015}.

Density estimators for $N$-body data have traditionally relied on depositing point
particles to a regular grid using cloud-in-cell, adaptive kernel smoothing, Voronoi tessellation, 
or some other such technique (see e.g. \citealt{Hoc1988}, \citealt{Ney2008}). These are subject to sampling noise 
and most do not give a well-defined density field
everywhere in space. This is especially true in regions of low particle density (voids), so we require a
smoother and more physically faithful method.

The approach of \citet{Abel2012} and \citet{Sha2012} instead discretizes the dark matter mass into
tetrahedra, where the particles are demoted to Lagrangian tracers that serve as
vertices of these tetrahedral mass elements. In more technical terms, we model the cold dark matter fluid as
a piecewise-linear approximation to a three-manifold embedded in the six-dimensional phase-space. 
This approach gives a density field that is well-defined everywhere in configuration space, in contrast with a set of
point particles as in the traditional $N$-body case. So far, it has been applied successfully to
gravitational lensing \citep{Ang2014}, large-scale structure \citep{Hah2015}, 
simulations \citep{Hah2013, Hah2016, Sou2015}, and visualization \citep{Kae2012}.

We apply a piecewise-linear density interpolation across each tetrahedron. This gives the smoothest
density maps and has proven the most effective at revealing voids through the watershed transform.
The density at each tracer vertex $v$ is given by a volume-weighted average of the masses of the
tetrahedra sharing that vertex, $t \in v$:
$$
\rho_v = \frac{\sum_{t \in v} m_t}{\sum_{t \in v} V_t}.
$$
This gives a density map that is smooth and continuous.

Given this discretization of the dark matter into a set of tetrahedra with linearly-varying density, 
we require a way of depositing
this density field on to a uniform Cartesian grid on which to calculate the watershed transform. We
do this using the conservative voxelization method of \citet{Pow2015}, which robustly calculates the exact
volume and moments of intersection between each tetrahedral element and each cubical cell in
the target grid, then computes the amount of mass to be transferred. This guarantees
that mass is globally and locally conserved and that the resulting density field is smooth and naturally
anti-aliased.

The final result is a density field on a Cartesian grid that is smooth and well-defined everywhere,
especially in voids where a particle sampling method would be insufficient.

\subsection{Watershed transform}

The watershed transform divides space into contiguous zones, the so-called watershed basins. Every watershed basin 
contains a local density minimum and it is delineated by a surface which is a 3D analogue of a ridge line in a 2D density 
field \citep[for a more detailed description see][]{Pla2007}. Watershed basins are typically much smaller than physical 
voids; therefore, the final step of the void finder involves merging basins into voids. Following the algorithm 
implemented in ZOBOV, we build voids by merging all adjacent watershed basins which are shallower 
than the initial one. Basins are merged according to the order of saddle points at the boundaries between them. 
The process of merging stops when it encounters a basin with a minimum density lower than the minimum density of the initial basin 
\citep[for more details see][]{Ney2008}. 

The way the adopted method builds voids is fully determined by orders of density minima and saddle points of the density field. 
The total volume of all voids exceeds the total volume of watershed basins (volume of the simulation box). This means that voids 
overlap in space and form a hierarchical network such that every void is a subvoid of only one void and it may 
contain an unlimited number of subvoids \citep[see e.g.][]{Lav2012}. The network consists of many levels of hierarchical 
relations between voids. Hereafter we refer to a void-subvoid sequence as, respectively, the top and bottom levels of the void hierarchy.

We perform watershed transform and build voids using the \textit{jozov} module of ZOBOV. Despite the fact that the ZOBOV computes 
density using Voronoi tessellation, the \textit{jozov} module can easily operate on different density grids. This only requires 
appropriate labelling pixels and specifying connectivity between them. We assume that every pixel of a regular cubical 
grid has $26$ adjacent pixels sharing its faces or vertices.

\begin{figure*}
\centering
\begin{tabular}{c}
\includegraphics[width=0.98\textwidth]{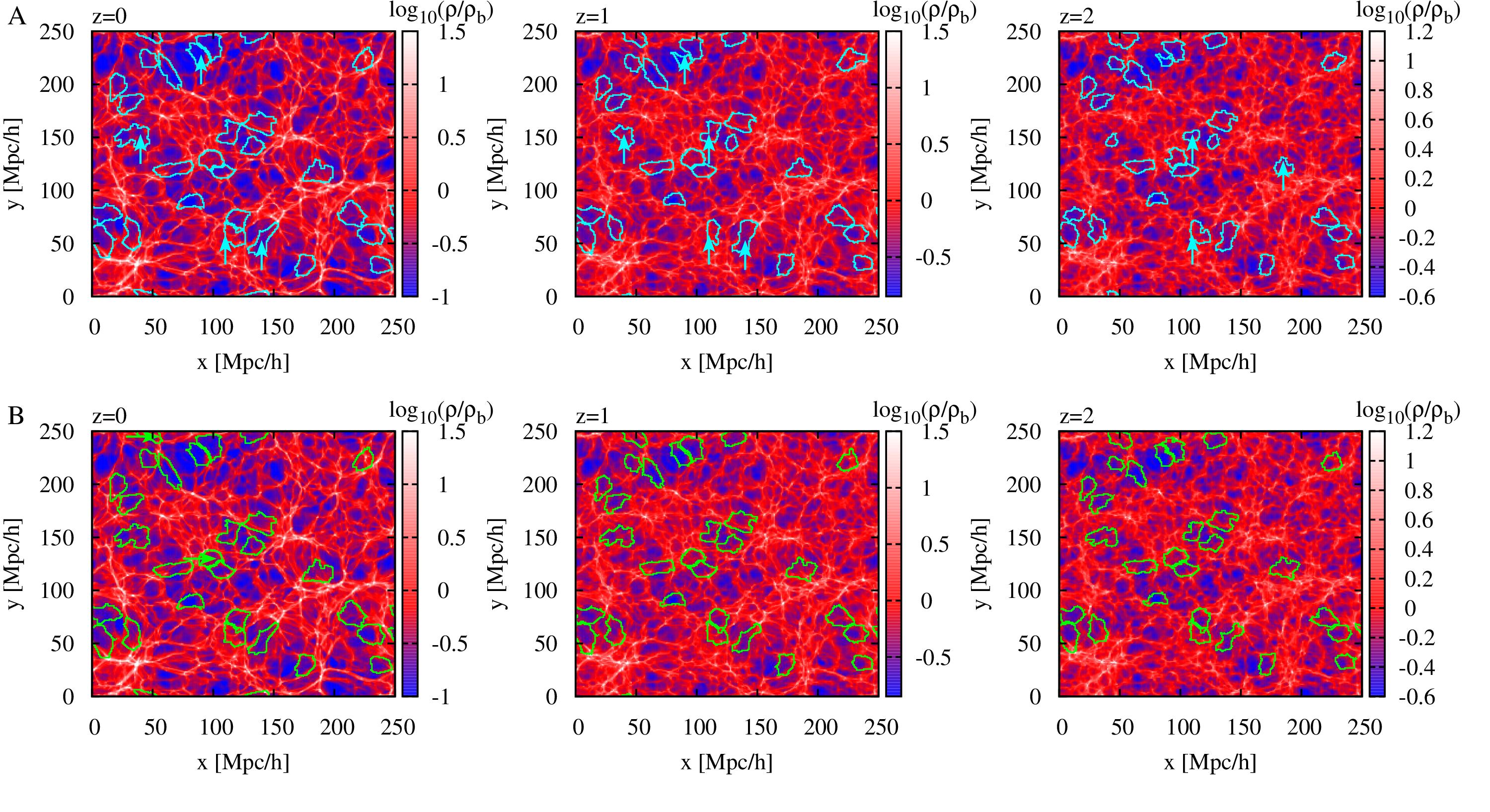} \\
\hline
\includegraphics[width=0.98\textwidth]{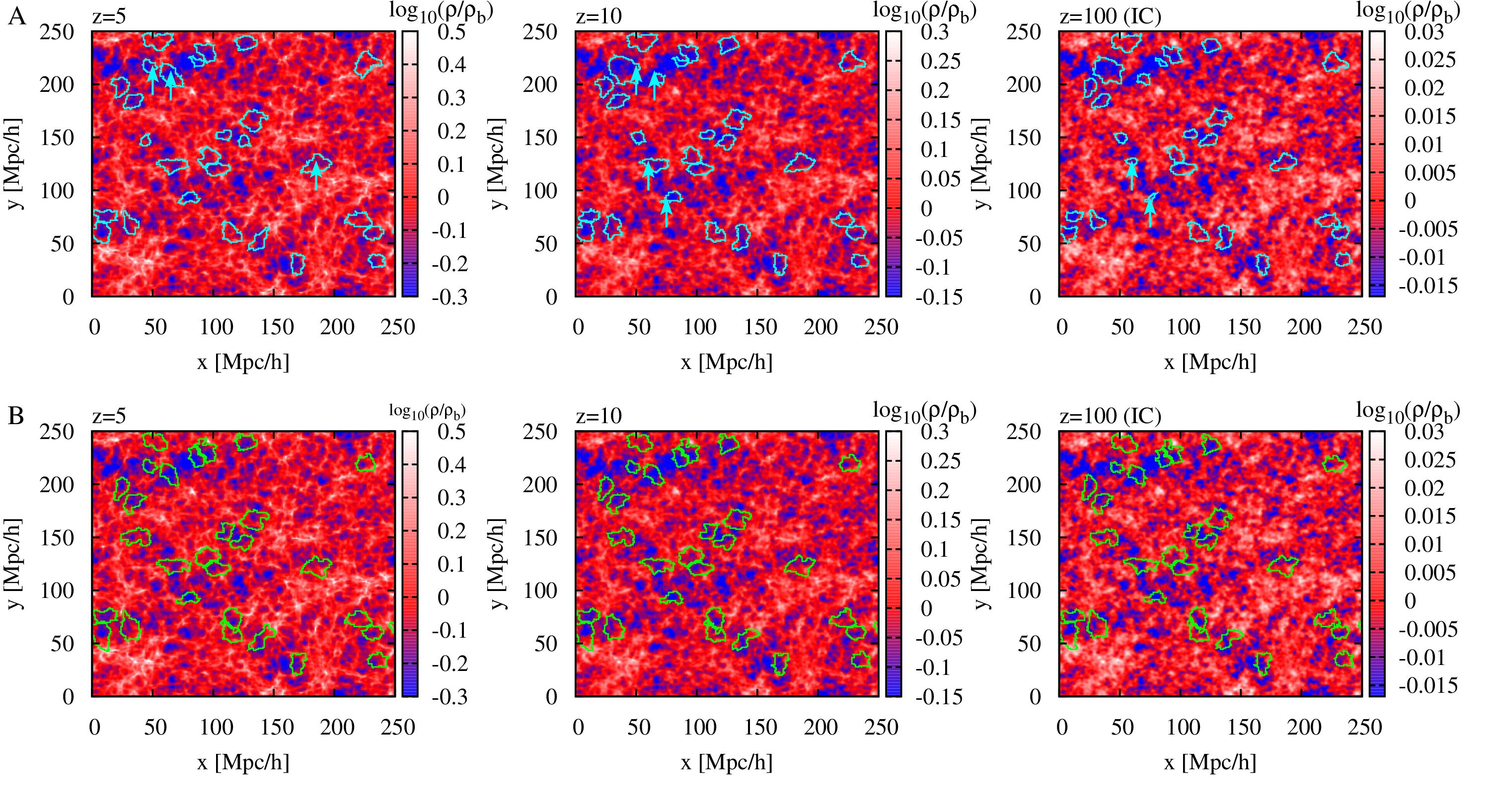}\\
\end{tabular}
\caption{Two methods of tracking voids in cosmological simulations. 
The upper row (A) shows evolution of voids found by matching 
voids between adjacent snapshots (with watershed basins merged into voids 
in every snapshot independently). The bottom row (B) shows evolution of voids found by 
means of tracking watershed basins 
back in time under the assumption that voids consist of the same groups of basins 
at all times (with watershed basins merged into voids only once at $z=0$). 
Both methods start with the same group of voids 
at $z=0$ and track them back in time to initial conditions. 
Every panel shows voids boundaries projected on to 
the image plane and log of the density field inside a $10\hMpc$ thick slab. Voids at $z=0$ 
(upper left panels) have effective radii between $8\hMpc$ and $10\hMpc$ and lie no further 
than $30$ per cent of their effective radii from the image plane. Results of void tracking 
are presented in the consecutive panels corresponding to snapshots at redshifts $z=1,\;2,\;5,\;10,\;100\,\textrm{(initial conditions)}$. 
Tracking watershed basins (row B) yields evolutionary tracks of voids which are continuous 
at all times (voids coevolve with their boundaries). On the other hand, matching voids between adjacent 
snapshots (row A) gives rise to spurious reconfigurations resulting from reordering of density minima 
and saddle points, in large part due to unresolved small-scale modes and numerical rounding. 
Voids undergoing most visible reconfigurations are indicated with arrows (two snapshots 
before and after the event).
}
\label{maps}
\end{figure*}

In most applications, void finders invoke density thresholds which define the lowest density 
on the boundary surface between every two adjacent voids (ZOBOV) or impose a limit on density in void 
cores. Since our main goal is to trace evolution of voids from initial conditions to the present time, 
our void finder cannot rely on any density threshold. We therefore do not assume any density limits 
on void cores and void barriers and find voids in a fully non-parametric way. This approach is feasible in our case because 
the adopted density grid smoothens the density on scales 
comparable to dark matter haloes and thus it automatically eliminates the possibility of spurious detections in high 
density regions \citep[][]{Ney2008}. It also brings back the original idea of ZOBOV devised as a non-parametric void finder.

\subsection{Cosmological simulation}

For the purpose of our studies we ran a low resolution $N$-body simulation 
of dark matter structure formation in a $\Lambda$CDM cosmological model 
with Planck cosmological parameters, i.e. $\Omega_{\rm m}=0.31$, $\Omega_{\Lambda}=0.69$, $\sigma_{8}=0.82$ and 
$h=0.67$ \citep{Planck2014a}. The initial conditions were generated using the Multi-Scale Initial Conditions code 
\citep[MUSIC; ][]{Hahn2011} with the transfer function given by a fitting formula obtained by \citet{Eis1998}. The simulation 
box has a side length equal to $250\hMpc$ and contains $N_{\rm p}=256^{3}$ particles. The adopted mass resolution allows 
us to resolve voids with effective radii larger than $\sim 5\hMpc$. 

The simulation was run using the Gadget-2 $N$-body code \citep{Spr2005} and snapshots used for tracing void evolution 
were saved with a temporal resolution of $\Delta a/a=0.05$. The density field used for finding voids in all 
snapshots was computed on a regular grid with $N_{\rm g}=256^{3}$ cells corresponding to a grid spacing of 
nearly $1\hMpc$. The adopted grid size smoothens the density field on scales comparable to largest 
dark matter haloes.

\section{Void tracking}

Evolution of voids is distinct from evolution of dark matter haloes in many respects. Many evolutionary processes do not 
have their counterparts in mechanisms of halo formation. Well known examples of such processes are disappearance 
of voids caused by collapse of walls around them or void merging due to dilution of boundaries between them. The 
comprehensive description of all these processes as well as theoretical framework based on the excursion set theory was 
outlined by \citet{She2004}. In our work we shall address just one aspect of this complexity of voids evolution, namely 
evolution of various properties of voids between initial conditions and redshift $z=0$. A practical problem arising 
here is a robust method for tracking voids in cosmological simulations.

\subsection{Simple void matching}

Voids can be tracked in simulations in a similar way as dark matter haloes. A common tool devised for this purpose is a merger 
tree. In the context of dark matter haloes, merger trees represent physical processes of halo formation and tracking dark matter 
haloes is reduced to finding the main progenitors. From a more abstract point of view, merger tree is merely a means of linking 
or associating objects detected at different snapshots. It is therefore tempting to think that the same algorithms used for 
building halo merger trees can be utilized to track cosmic voids by finding a sequence of void progenitors. 
This approach was already presented by \citet{Sut2014c}. 
Their method employed the VELOCIrapter tree builder 
code, a publicly available code for generating halo merger trees, which was directly applied to catalogues of voids 
found in a cosmological simulation. This direct implementation of the halo merger tree code was feasible thanks 
to the fact that voids were found on a Voronoi grid and therefore they were defined by IDs of particles. On the one hand, 
one can appreciate that the same merger tree code can be applied both to haloes and voids. On the other hand, one 
can notice the following caveat of this approach. As pointed out by \citet{Sut2014c}, identifying voids 
by particles is subject to some degree of ambiguity. Most particles occupy boundaries of voids and therefore they are often 
exchanged between adjacent voids. This makes identification of voids in different snapshots less accurate. The process 
of matching voids between different snapshots can be substantially improved if voids were defined in terms of comoving 
volume they occupy (IDs of pixels of a regular grid in comoving space) rather than mass they contain (IDs of particles). 
This can be automatically achieved by running a void finder on density fields computed on a regular grid adopted 
in our density estimator.

We use our void finder described in section 2 to detect voids in all snapshots of the simulation. Voids are specified 
as contiguous groups of pixels. We track every individual void by creating a sequence of the best matching voids 
between successive pairs of adjacent snapshots. The best matching void is found by maximizing degree of overlap 
in space which is quantified by $n_{AB}^{2}/(n_{A}n_{B})$, where $n_{AB}$ is the number of pixels shared by 
void from snapshot $A$ and void from snapshot $B$, $n_{A}$ and $n_{B}$ are the numbers of pixels in the two voids. 

\begin{figure}
\centering
\includegraphics[width=0.48\textwidth]{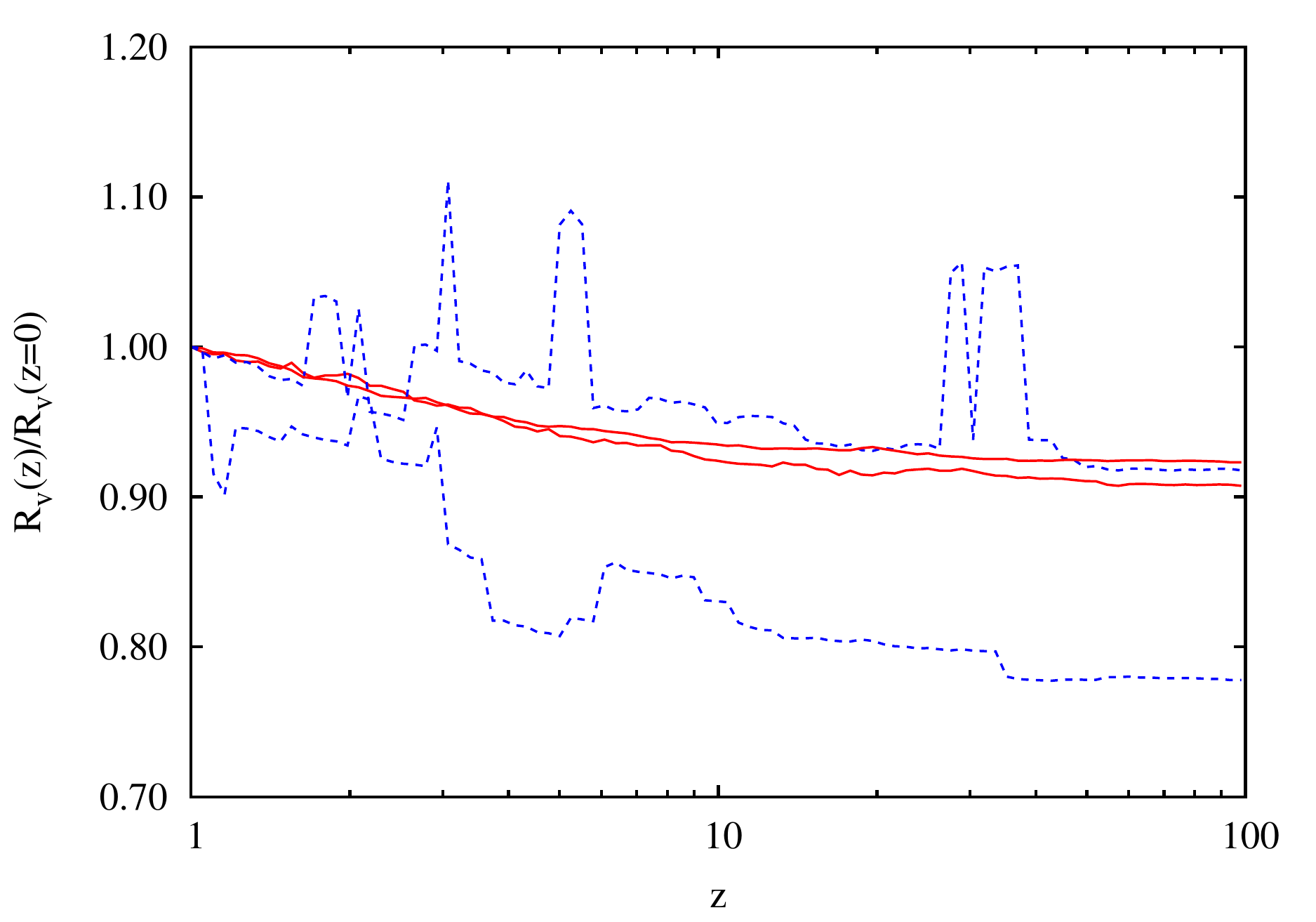}
\caption{Redshift evolution of effective radii of two typical voids selected from $z=0$ snapshot shown 
in Fig.~\ref{maps} (indicated by arrows on $z=0$ panel in row B ). The solid red lines show results from tracking 
watershed basins of $z=0$ voids back in time (illustrated by panels B in Fig.~\ref{maps}) and 
the blue dashed lines from matching voids between 
successive pairs of adjacent snapshots (illustrated by panels A in Fig.~\ref{maps}). The former method 
yields a continuous evolution of void boundaries over time resulting in smooth redshift profiles. The 
latter method generates spurious and instantaneous reconfigurations, well visible as strong 
discontinuities of the profiles.
}
\label{Rv-comparison}
\end{figure}

The upper panels (A) in Fig.~\ref{maps} show an example of tracking a group of voids from redshift $z=0$ to 
initial conditions. The contours 
show projections of voids on to 
the image plane and the map shows log of the density integrated inside a slab of a $10\hMpc$ thickness. In order to 
maximize correspondence between the density map and voids we select voids which lie no further than $30$ per cent 
of their effective radii from the image plane at $z=0$, where the effective void radius $R_{\rm v}$ is radius of a sphere 
enclosing volume $V$ equal to the void's volume, i.e.
\begin{equation}
R_{\rm v}=\Big(\frac{3V}{4\pi}\Big)^{1/3}.
\end{equation}
All voids shown in Fig.~\ref{maps} have effective radii in a range between $8\hMpc$ and $10\hMpc$ at $z=0$, well 
above the resolution limit.

Fig.~\ref{maps} illustrates that the method of matching voids between adjacent snapshots yields in principle a 
reasonable mapping between void boundaries at different redshifts. Some voids clearly coevolve with the cosmic web and remain 
delineated by the same boundary walls at all redshifts. However, a closer inspection of the plot reveals a group of voids 
which undergo instantaneous reconfigurations such as loss of volume or incorporation of some parts of adjacent voids 
(see voids indicated by arrows). 
These processes result merely from reordering of density minima and saddle points which determine 
a way how watershed basins are merged into voids. Therefore, they are related to the adopted definition of voids 
rather than genuine evolution of voids. The problem becomes even more complicated if we realize that 
reordering of critical points in the density field is caused in some part by unresolved small-scale modes 
and thus it is a stochastic process.  Another important factor is numerical rounding.

We assess the impact of numerical resolution on the apparent reconfigurations of voids by comparing voids found in two 
simulation runs with the same initial conditions but different mass resolutions. For this purpose we reran our main simulation with 
$N_{\rm p}=512^{3}$ particles, computed the density field using data downsampled to $N_{\rm p}=256^{3}$ 
and then found voids in the same way as described in section 2. Matching 
voids between the catalogues from high and low resolution runs we find that typical degree of matching is much smaller than for 
a naive one-to-one correspondence between both void catalogues. The mean space overlap $n_{AB}^{2}/(n_{A}n_{B})$ 
between best matched voids with effective radii between $6\hMpc$ and $20\hMpc$ is around $0.8$ (up to $20$ per cent relative 
difference in volume).

Fig.~\ref{Rv-comparison} illustrates more explicitly a stochastic nature of instantaneous reconfigurations of voids tracked 
by means of matching voids between successive pairs of adjacent snapshots. The dashed lines show the redshift evolution 
of effective radii of two representative voids from Fig.~\ref{maps} indicated by arrows on $z=0$ panel in row B. The redshift 
profiles clearly demonstrate that the reconfigurations occur in a random manner. In extreme cases, voids undergo 
a substantial change of their volumes and then return to their initial boundaries in the following snapshot. We conclude 
that a robust method for tracing evolution of voids and their properties should be capable of eliminating this purely numerical 
effect.

\subsection{Tracking watershed basins}

Spurious reconfigurations of voids demonstrated in the previous section result in large part from random reordering of the density 
minima and saddle points, caused by unresolved small-scale modes and numerical rounding. A simple way to circumvent this problem 
is to avoid merging watershed basins into voids at every snapshot independently. Instead, we propose to generate a void catalogue only once 
and then to track watershed basins back in time in order to find the corresponding voids at different redshifts. We therefore assume 
that voids consist of the same groups of watershed basins at all times. We expect that this assumption guarantees a continuous 
mapping between void interiors or boundaries over time. On the other hand, one disadvantage of this approach is 
the fact that it is not expected to tackle such processes as void destructions, which changes the number of voids over time, and 
voids mergers, which are associated with physical reconfigurations in void network. The former processes can be captured by 
finding voids at high redshift and tracking them forward in time.

We track watershed basins of voids from $z=0$ snapshot in a similar way as voids in the previous section. First we apply watershed 
transform to all snapshots. Then we match basins between every two adjacent snapshots in such a way that every basin 
contains the minimum of its best matching basin from the other snapshot. Watershed basins may not evolve in a continuous way. They can 
merge (more common) or split (less common) in the course of their evolution. Both processes are easily tracked as 
bifurcations in the linking relations between basins. For example, merging process occurs when at least two adjacent basins 
are lined to only one basin in a later time snapshot.

The bottom panels (B) in Fig.~\ref{maps} show evolution of voids found by tracking watershed basins back 
in time. For the sake of objective comparison with the previous method (see 
the corresponding panels in the upper row, A), we use the same group of voids selected 
from $z=0$ snapshot. Fig.~\ref{maps} clearly demonstrates how differences between voids found by the 
two methods gradually increase at high redshifts. Whereas many voids from the method based on matching 
voids between snapshots (panels B) undergo unphysical reconfigurations (instantaneous loss of volume 
or incorporation of adjacent space), voids found by tracking watershed basins (panels A) are consistently delineated 
by the same boundaries at all redshifts. Closer examination of how the 
contours evolve between the panels shows that the method  based on tracking watershed basins allows us 
to create a continuous mapping between voids interiors or boundaries at different redshifts.

Instantaneous reconfigurations in void tracking by means of matching voids between adjacent 
snapshots occur quite often. $50$ per cent of voids change their volumes between adjacent snapshots by more 
than $50$ per cent and nearly $70$ per cent of them by more than $30$ per cent. Reconfigurations may happen 
more than once and are more frequent at low redshifts. The corresponding fractions of voids undergoing comparable 
volume evolution are negligible ($\sim10^{-3}$) when voids are tracked using the new method. This is illustrated 
by Fig.~\ref{Rv-comparison} which compares redshift evolution of effective radii of two representative voids 
selected from $z=0$ panel in Fig.~\ref{maps} (voids indicated with arrows on $z=0$ panel in row A) and tracked back 
in time using the new method (solid red lines) or the previous method based on matching voids between adjacent snapshots (dashed blue lines). 
It is clear that the method based on tracking watershed basins eliminates spurious effects of the previous one and yields continuous 
redshift profiles.

\section{Evolution of voids}

Here we use our new algorithm for tracking voids in cosmological simulations to study evolution of their basic properties. 
Since our main focus is to compare mature voids to their early state in initial conditions, it is natural to create a void catalogue 
at redshift $z=0$ and then to find the corresponding voids in all remaining snapshots at higher redshifts. For practical reasons, 
we restrict our study to around $6100$ voids with effective radii selected from a narrow range between $6\hMpc$ and $20\hMpc$ at $z=0$. 
We therefore neglect a class of voids which might have collapsed before that redshift \citep{She2004} and all results presented 
in this section show evolution of merely voids existing at redshift $z=0$.

\begin{figure}
\centering
\includegraphics[width=0.5\textwidth]{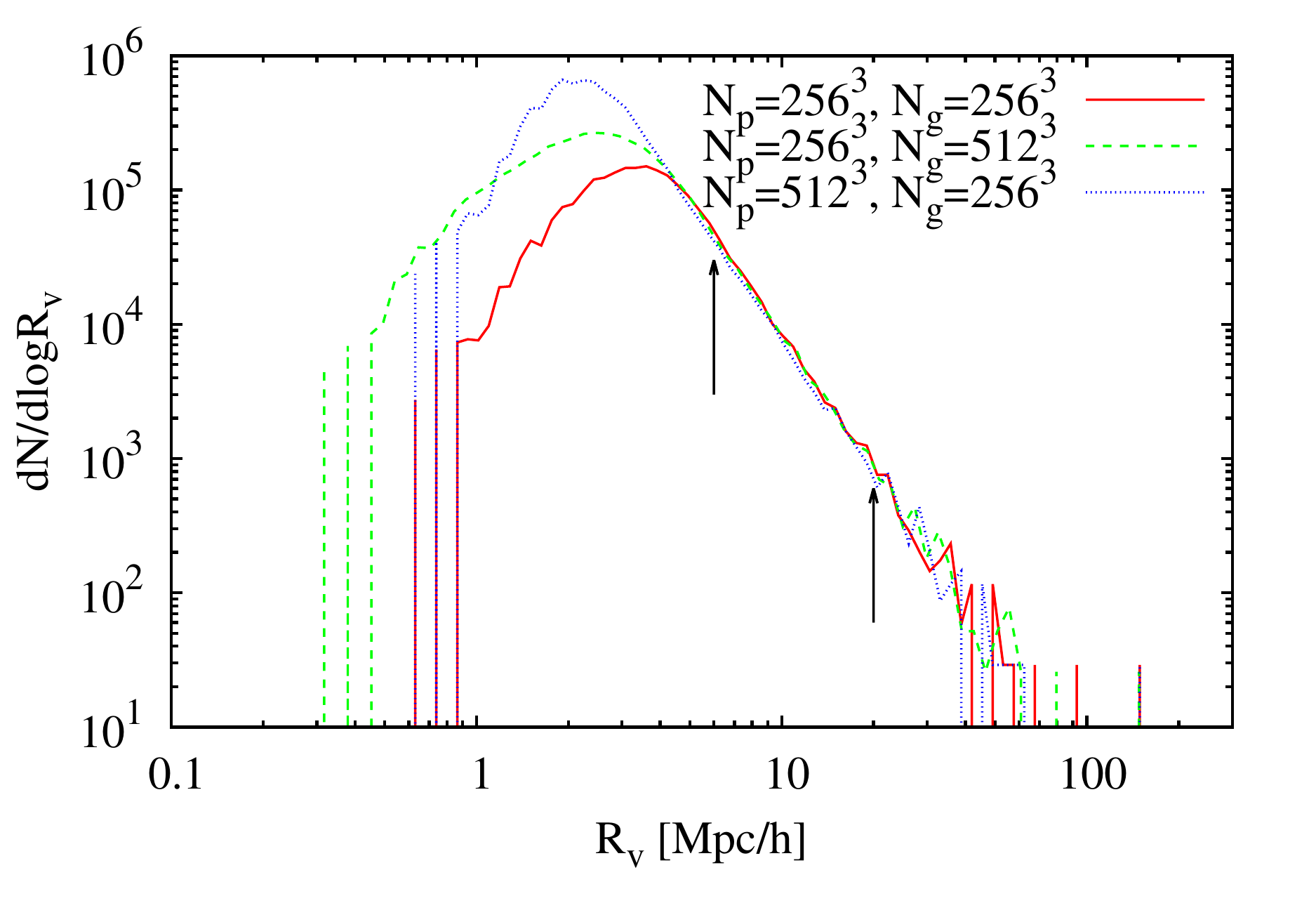}
\caption{Distribution of void effective radii at redshift $z=0$ in simulations with three different mass or 
grid resolutions given by the number of particles, $N_{\rm p}^{3}$, and cells of the density grid, $N_{\rm g}^{3}$, 
used for void finding. The black arrows show the range of void sizes adopted in this study. The selected voids have sizes well above the 
resolution limit.}
\label{pdf-sizes}
\end{figure}

The selected voids have sizes comparable to typical sizes of voids found in observations \citep{Paz2013}. They are also well above the 
resolution limit of our primary simulation with $N_{\rm p}=256^{3}$ particles 
and $N_{\rm g}=256^{3}$ cells of the density grid used for finding voids. This is illustrated by Fig.~\ref{pdf-sizes} which shows the distribution of 
void effective radii in simulations with an increased mass or density grid resolution. For the adopted density grid resolution, 
voids contain from $10^{3}$ to $4\times10^{4}$ pixels, arguably sufficient for a robust computation of all void properties 
considered in this work. For the sake of a sanity check, in a few instances we plot results from an analysis based 
on an increased grid resolution with $N_{\rm g}=512^{3}$ cells in order to demonstrate numerical convergence.

\begin{figure}
\centering
\includegraphics[width=0.49\textwidth]{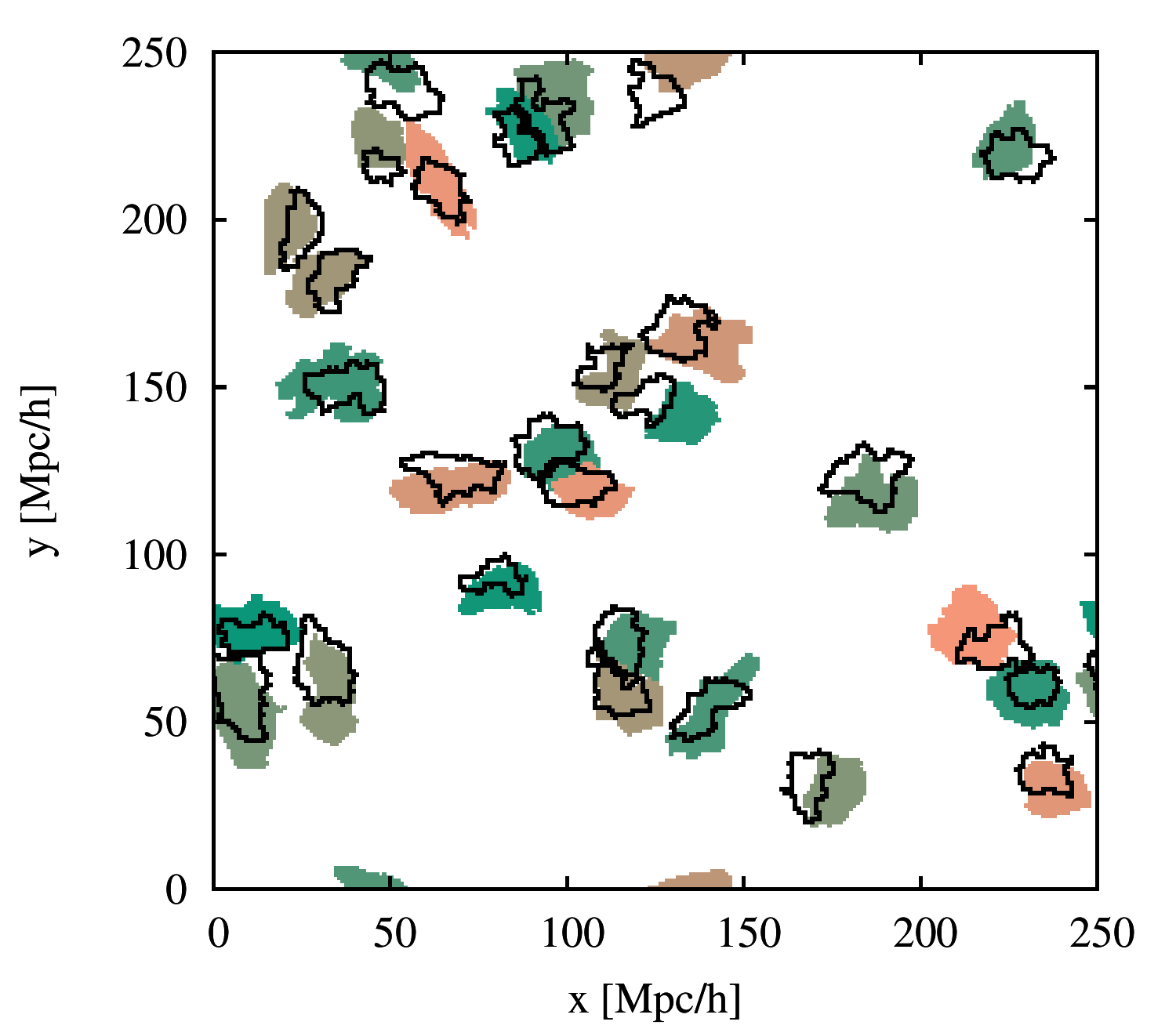}
\caption{Projected boundaries of voids from Fig.~\ref{maps} at redshift $z=0$ (filled contours) and 
in initial conditions at $z=100$ (maximally overlapping black contours). Voids at different redshifts are found 
by tracking watershed basins of $z=0$ back in time. The picture illustrates a variety of evolutionary processes: 
bulk motion, anisotropic expansion or contraction leading to evolution of shapes and orientation of principle axes.
}
\label{void-proto}
\end{figure}

Fig.~\ref{void-proto} illustrates variety of processes transforming initial voids into well developed voids at redshift $z=0$. 
In this figure, we compare projected boundaries of voids from Fig.~\ref{maps} at redshift $z=0$ (filled contours) and 
to the corresponding boundaries in initial conditions (black contours). The main agent distorting shapes 
of voids and changing their orientations are tidal forces. Depending on how uniform the gravitational acceleration is across 
the void interiors, some voids gain bulk velocities and move in space. Many voids undergo expansion driven 
by a super-Hubble flow of their host voids (compare with the density field in Fig.~\ref{maps}). Finally, although 
not displayed in Fig.~\ref{void-proto}, it is important to note that all voids evacuate matter from their cores to their 
boundaries decreasing densities in their central parts from $\sim\rho_{\rm b}$ in initial conditions to $\lesssim0.2\rho_{\rm b}$ at 
redshift $z=0$.

\begin{figure}
\centering
\includegraphics[width=0.49\textwidth]{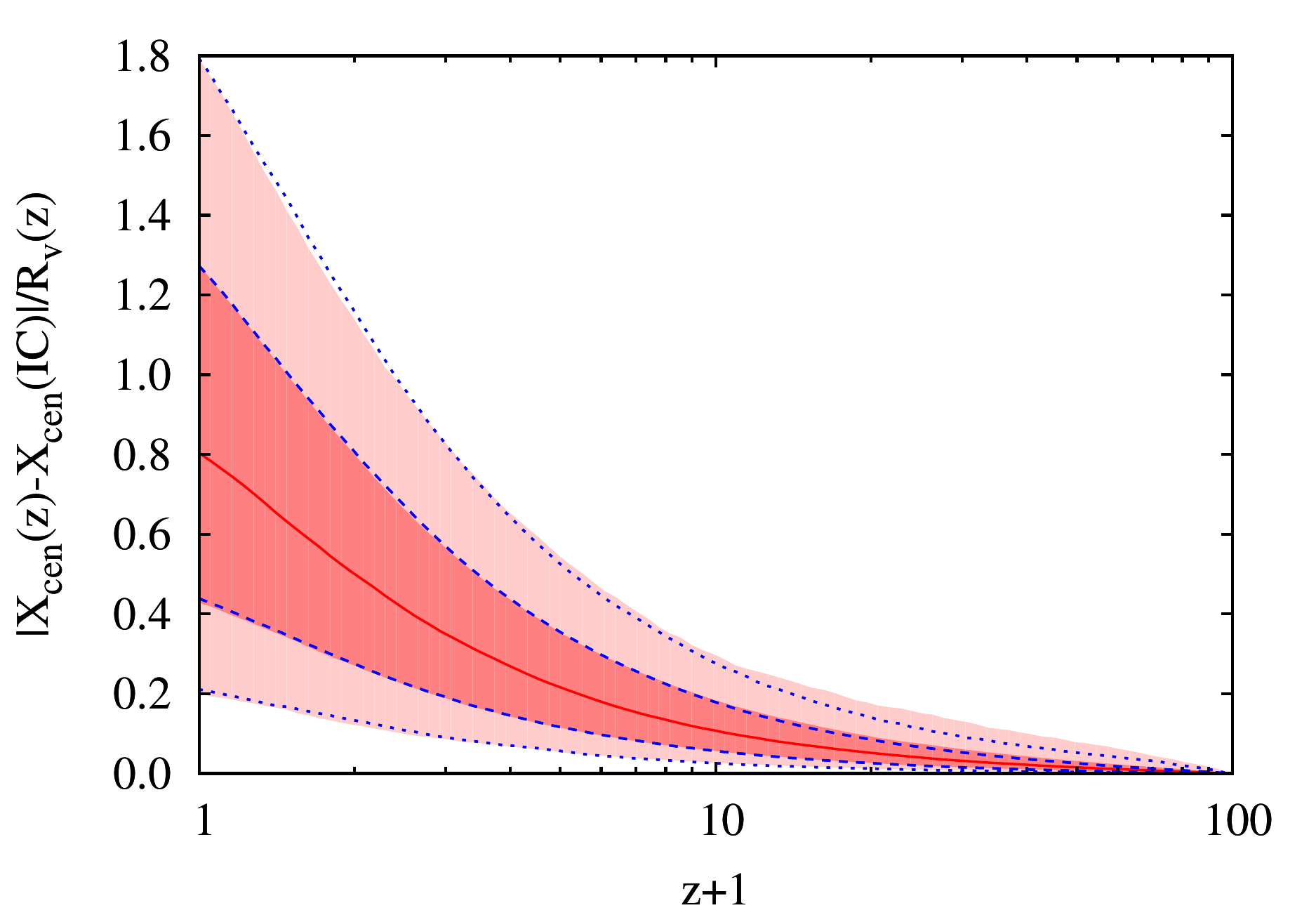}
\caption{Displacement of geometric centres of voids between initial conditions ($z=100$) 
and redshift $z$. Voids are traced in the simulation by means of tracking 
watershed basins of $z=0$ voids back in time. 
The shaded contours show the $1\sigma$ and $2\sigma$ ranges of 
the distribution and the solid red line is the median profile. The selected 
voids have effective radii $6\hMpc<R_{\rm v}<20\hMpc$ at $z=0$. The geometric 
centres move due to bulk motions of voids and anisotropic deformations of their boundaries. 
The dashed and dotted lines show the $1\sigma$ and $2\sigma$ ranges from 
an analysis based on an increased density grid resolution with $N_{\rm g}=512^{3}$ cells.
}
\label{shift-z}
\end{figure}

As already illustrated in Fig.~\ref{void-proto}, voids undergo substantial modifications of their initial boundaries. Fig.~\ref{shift-z} 
demonstrates spatial scales of this process in a more quantitative way. The shaded contours show the distribution 
of distances travelled by geometric centres $X_{\rm cen}$ (average position of all void pixels) between their initial positions in voids at $z=100$ and 
positions in consecutive snapshots of the simulation. We checked that these displacements remain virtually the same 
when pixels are weighted by inverse density. This shows that the measured displacements of void geometric centres reflect 
genuine evolution of void interiors: bulk motions \citep{Lam2016} and anisotropic deformations by tidal forces. As naturally expected 
for a linear evolution, geometric 
centres of voids at high redshifts remain at rest. The bulk growth of distances travelled by geometric centres of voids 
occur at low redshifts. Typical displacements reach $80$ per cent of the final effective radii signifying that spatial scales 
of void evolution are comparable to their sizes. This in turn reflects a basic fact that typical sizes of void deformations 
are comparable to typical distances travelled by particles in cold dark matter cosmological model.

\subsection{Void sizes}

\begin{figure}
\centering
\includegraphics[width=0.49\textwidth]{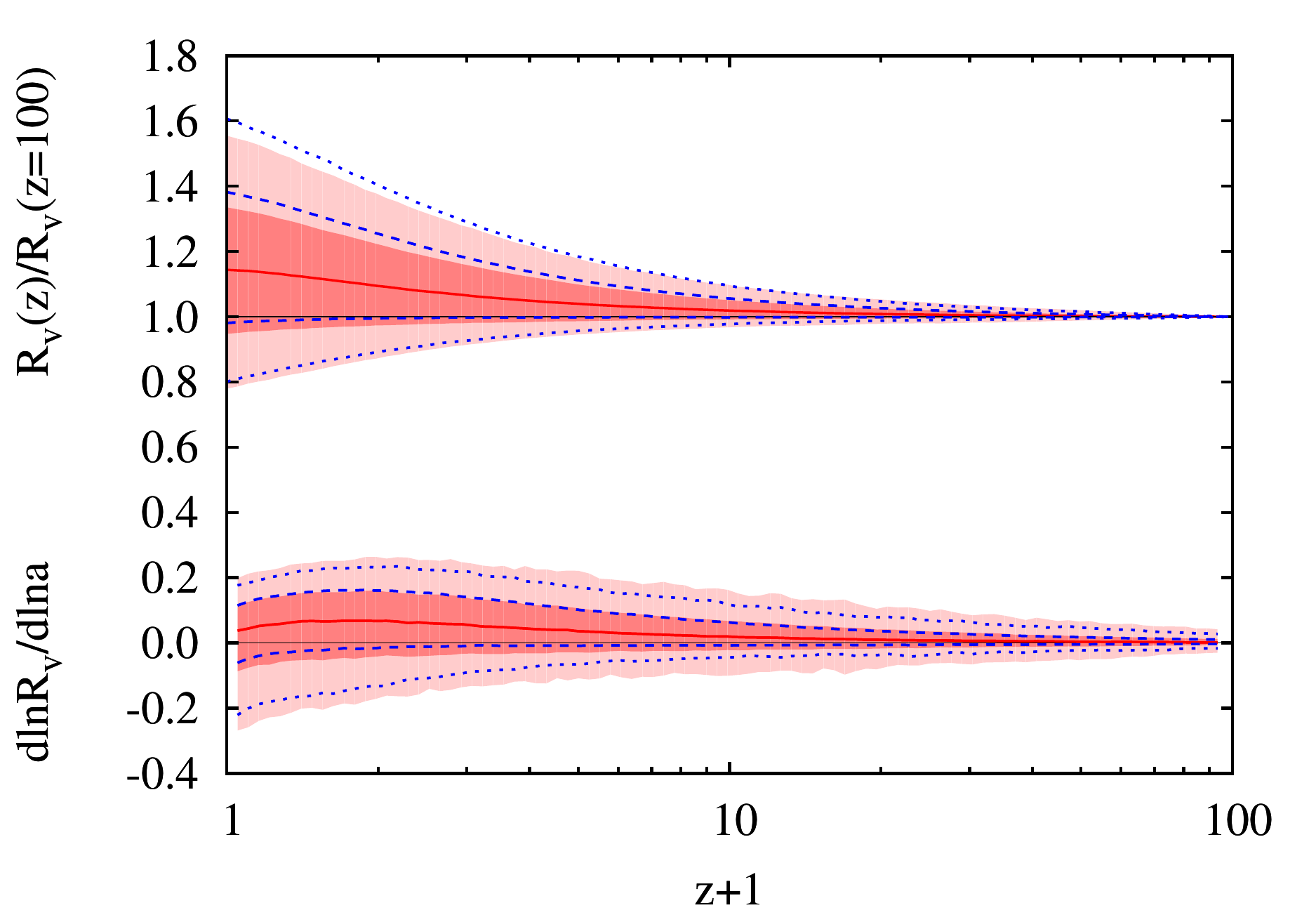}
\caption{Redshift evolution of void effective radii and their logarithmic growth. Voids are 
traced in the simulation by means of tracking watershed basins of $z=0$ voids back in time. 
The shaded contours show the $1\sigma$ and $2\sigma$ 
ranges of the distribution and the solid red line is the median profile. The selected 
voids have effective radii $6\hMpc<R_{\rm v}<20\hMpc$ at $z=0$. 
The dashed and dotted lines show the $1\sigma$ and $2\sigma$ ranges from 
an analysis based on an increased density grid resolution with $N_{\rm g}=512^{3}$ cells.
}
\label{rad-evolution}
\end{figure}

\begin{figure}
\centering
\includegraphics[width=0.49\textwidth]{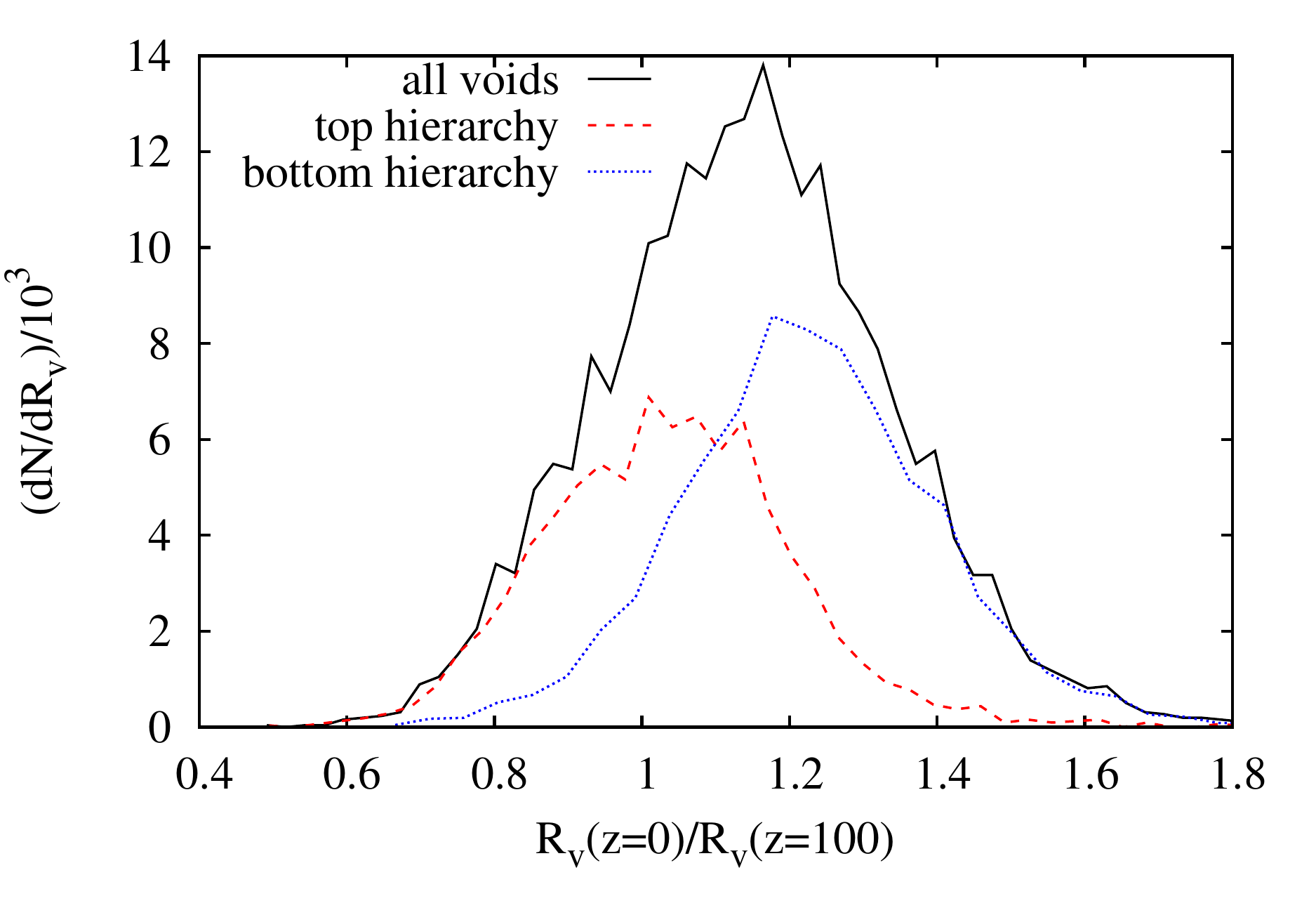}
\caption{Distribution of growths of void sizes from $z=100$ (initial conditions) 
to $z=0$. The dashed (red) and dotted (blue) lines show the distributions for voids from 
the two topmost levels of the void hierarchy and from the remaining lower 
levels (subvoids). The dynamics of voids reflects their positions in the void hierarchy: 
the most expanding voids (the upper tail in the overall distribution) lie at the bottom of the hierarchy 
and the most collapsing ones (the lower tail in the overall distribution) lie at the top of the hierarchy. 
The selected voids have effective radii $6\hMpc<R_{\rm v}<20\hMpc$ at $z=0$.}
\label{rad-hierarchy}
\end{figure}

Fig.~\ref{rad-evolution} shows effective radii of voids as a function of redshift. Most voids undergo slow expansion increasing 
their effective radii typically by $15$ per cent. The process of expansion is characteristic for voids embedded in larger underdense regions 
which accelerate local expansion rate with respect to the global Hubble flow (super-Hubble flow). The opposite process may occur 
if walls around voids are sufficiently massive to initiate collapse and thus contraction of the voids interiors. Nearly $20$ per cent of 
voids selected at redshift $z=0$ undergo this process. The asymmetry between the number of contracting and expanding 
voids reflects asymmetry between overdense and underdense regions developed in a phase of non-linear evolution. Underdense 
regions fill more space and therefore a void-in-underdensity configuration is more likely than void-in-overdensity.

Evolution of void sizes appear to be much less prominent than displacements of geometric centres shown in Fig.~\ref{shift-z}. Whereas 
there is a high probability that void retains its initial volume, its geometric centre is displaced by at least $20$ per cent of the final effective 
radius. We therefore expect that evolution of void boundaries is much more complex than volumes they enclose. 
In fact, it is easy to imagine a number of processes changing shape or roughness of the boundary surface and preserving the enclosed volume.

The shaded contours in the bottom part of Fig.~\ref{rad-evolution} show the distribution of the logarithmic growth of effective radii as a function 
of redshift. The growth at redshift $z(a)$ is estimated as 
\begin{equation}
\frac{\textrm{d}\ln R_{\rm v}}{\textrm{d}\ln a}=\frac{\ln[R_{\rm v}(a+\delta a)/R_{\rm v}(a)]}{\ln[(a+\delta a)/a]}
\end{equation}
and averaged over two adjacent snapshots, i.e. $\delta a=\pm\Delta a$. Our results point to a much slower growth than what was obtained 
in recent studies by \citet{Sut2014c}. Typical logarithmic growth of void effective radii in our work is around $0.1$ with a maximum at $z\approx 1$. 
We note that this estimate would substantially increase if the effect of spurious reconfigurations discussed in section 3 was not corrected.

The physical reason whether voids expand or contract is directly related to how much mass is retained in void boundaries. If the mass 
overdensity deposited in void boundary exceeds the mass underdensity in void, then walls around the void start to collapse 
and thus void itself starts to contract. Coincidentally, mass deposited in void boundary determines also its location in the void hierarchy: 
voids delineated by massive walls are more likely at the top levels of the hierarchy, whereas voids delineated by weak walls are more likely 
subvoids of larger voids (bottom of the hierarchy). We can therefore expect that void dynamics reflects its position in the void hierarchy. 
Fig.~\ref{rad-hierarchy} confirms this supposition. The figure shows the distributions of the growth of void sizes in different groups of 
voids split according to their positions in the void hierarchy: a half of voids at the top-most or bottom-most levels of the hierarchy. 
It is evident that all most contracting voids lie at the top of the hierarchy, while all most expanding voids belong to the bottom levels of the hierarchy 
(subvoids). This connection holds if we narrow the range of effective radii to $6\hMpc<R_{\rm v}<8\hMpc$. 
This indicates that the primary relationship between expansion rate of voids and their sizes 
(large voids tend to expand and small voids tend to collapse; see e.g. \citet[][]{Ham2014, Sut2014c}) is additionally modulated by 
their positions in void hierarchy.

\subsection{Shapes}

\begin{figure}
\centering
\includegraphics[width=0.49\textwidth]{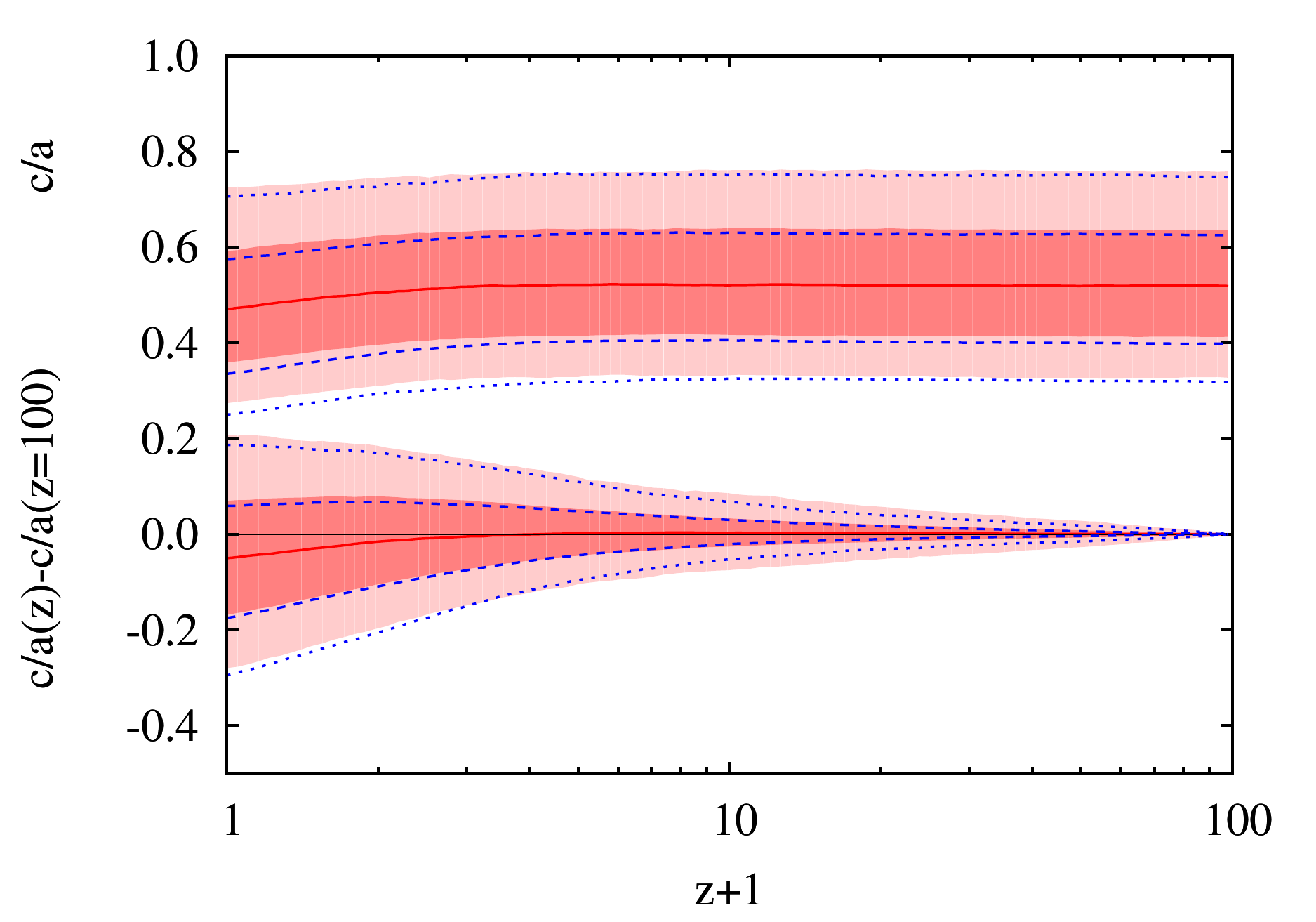}
\includegraphics[width=0.49\textwidth]{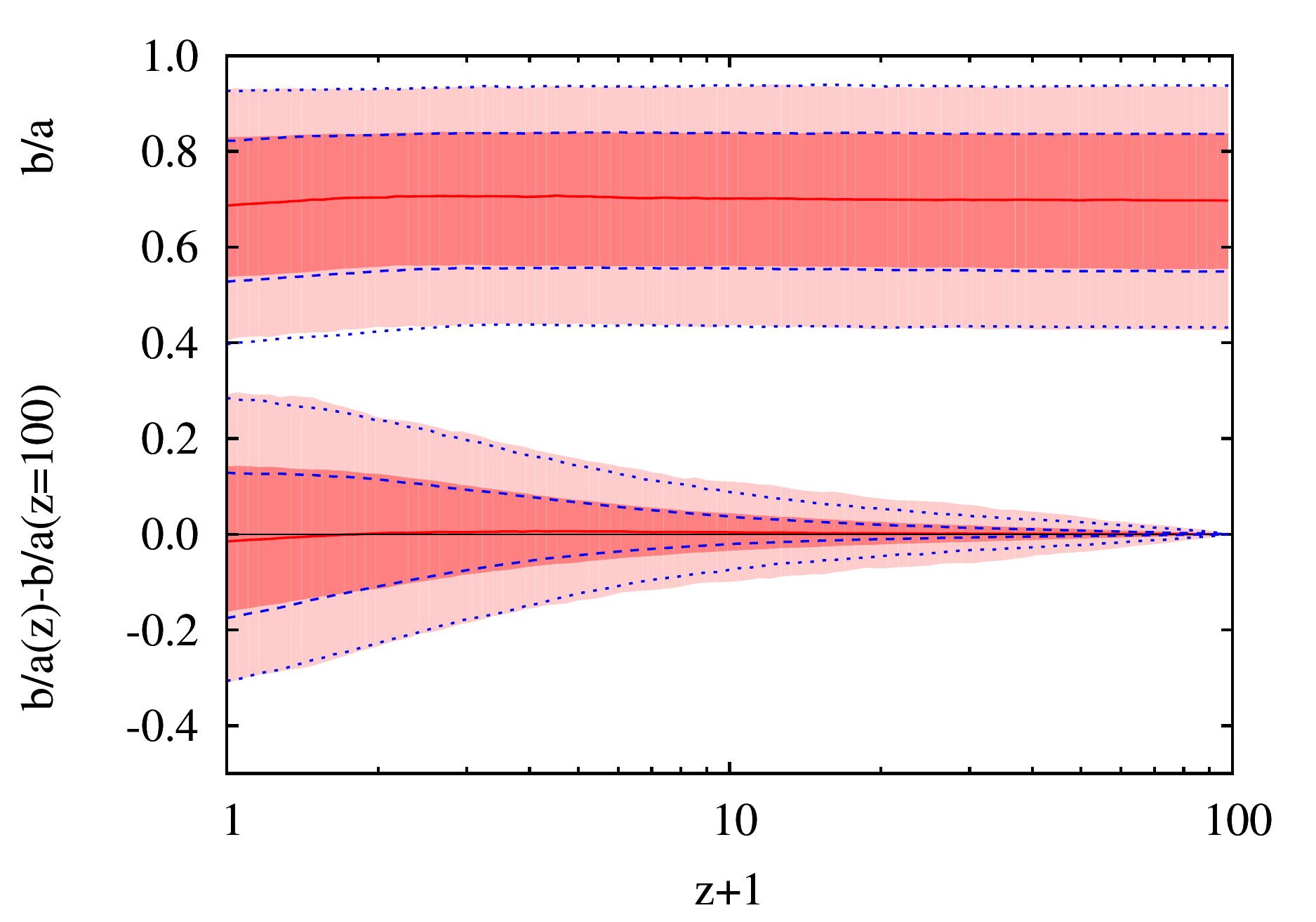}
\caption{Shapes of voids as a function of redshift 
(top panel: major-to-minor axis ratio of the shape ellipsoids,  $c/a$; bottom panel: 
medium-to-minor axis ratio of the shape ellipsoids, $b/c$). The shaded contours 
show the $1\sigma$ and $2\sigma$ ranges 
of the distribution at every redshift. The top contours in both panels show 
the overall distributions of axial ratios and the lower ones demonstrate evolution of shapes of 
individual voids relative to the initial shapes at $z=100$ (initial conditions). Individual 
voids change substantially their shapes; however, the overall distribution of axial ratios 
remains nearly the same. 
The dashed and dotted lines show the $1\sigma$ and $2\sigma$ ranges from 
an analysis based on an increased density grid resolution with $N_{\rm g}=512^{3}$ cells.
}
\label{shape-z}
\end{figure}

\begin{figure}
\centering
\includegraphics[width=0.49\textwidth]{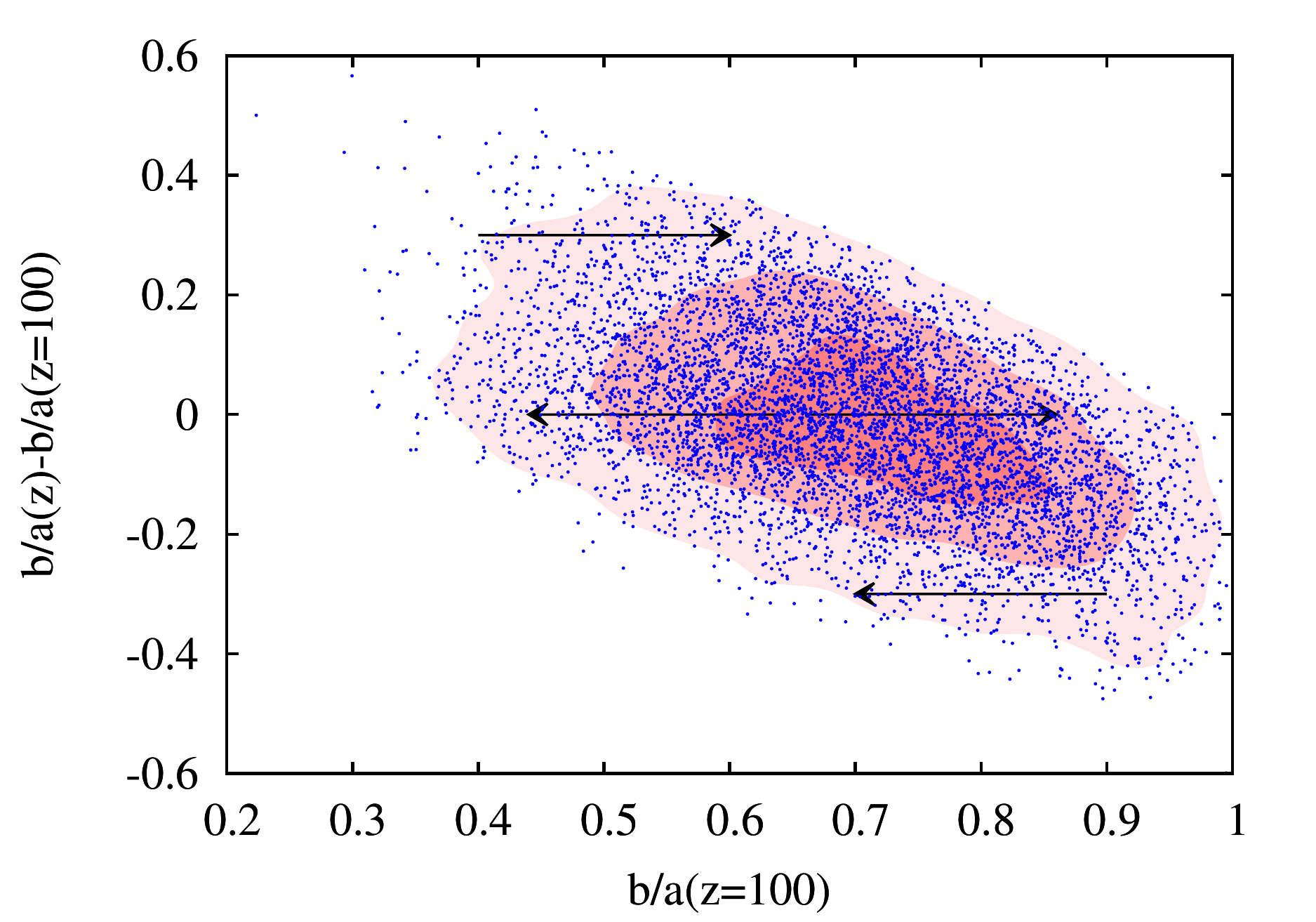}
\caption{Evolution of void medium-to-minor axis ratios  from $z=100$ (initial conditions) 
to $z=0$. The arrows show typical directions 
of the shape evolution in different parts of the initial shape distribution: most elongated 
($b/a\lesssim0.5$) and most spherical ($b/a\gtrsim 0.8$) voids evolve towards 
an average shape ($b/a\approx 0.65$), whereas voids with shapes around 
the average have equal chances of becoming more elongated or more spherical 
compensating loss in the tail of the shape distribution. This collective evolution of voids 
prevents the overall shape distribution from undergoing noticeable changes over time 
(see Fig.~\ref{shape-z}).
}
\label{db-b}
\end{figure}

Fig.~\ref{void-proto} shows evidently that shapes of voids evolve in time. Here we provide a more quantitative 
description of this process. In order to quantify shapes of voids, we compute the shape tensor $S$ given by
\begin{equation}
S_{ij}=\sum_{k=1}^{n} [e_{i}\cdot(r_{k}-X_{\rm cen})][e_{j}\cdot(r_{k}-X_{\rm cen})],
\label{shape-tensor}
\end{equation}
where $n$ is the number of pixels in void, $r_{k}$ is the position vector of $k$-th pixel, $X_{\rm cen}$ is the position vector of 
void geometric centre (average position of all void pixels) and $e_{i}$ with $i=1,2,3$ are unit vectors oriented along $x$, $y$ and $z$ axis of the simulation box. 
Diagonalizing the shape tensor we obtain its eigenvalues: $s_{a}$, $s_{b}<s_{a}$ and $s_{c}<s_{b}$, which are proportional to squares of 
the principle axes of an ellipsoid approximating void shape. We measure void shapes in terms of axial ratios of the shape ellipsoids, i.e. 
major-to-minor axis ratio $a/c=\sqrt{s_{a}/s_{c}}$ and medium-to-minor axis ratio $b/c=\sqrt{s_{b}/s_{c}}$.

Fig.~\ref{shape-z} shows axial ratios $a/c$ (top panel) and $b/c$ (bottom panel) of voids as a function of redshift. 
The upper contours on every panel show the overall distributions of axial ratios, while the lower contours show 
the distributions of a relative change in axial rations of all individual voids, between redshift $z$ and initial conditions. 
Comparing our results with those available in the literature (for $z=0$ only), we find remarkable agreement regardless of 
differences in the adopted definitions of voids \citep[see e.g.][]{Pla2008, Sha2006}.

As expected from the linear perturbation theory, voids at high redshift do not evolve their boundaries and thus retain their shapes. 
The bulk change of the shapes occurs at later time when every individual void can increase or decrease its axial ratio by up to 
$\sim0.2$, which is comparable to intrinsic scatter in the overall distributions. For the most typical evolutionary paths of voids, 
the $b/c$ axial ratio does not change and the $c/a$ axis ratio marginally decreases.

It is quite surprising to notice that a strong evolution of individual voids barely affects the overall distributions of axial ratios. 
One could expect that evolution of individual voids would at least broaden the distributions of axial ratios. However, what 
is clearly readable from Fig.~\ref{shape-z} is that the distribution of $b/c$ axis ratios remains the same at all redshifts, whereas 
the $a/c$ distribution exhibits only a small shift towards smaller values. The reason why the overall distributions of axial ratios do not 
evolve over time is related to phenomenon of collective evolution of voids. It is explained in Fig.~\ref{db-b} which 
shows the total change in $b/c$ axial ratio as a function of their initial values 
calculated for voids found in initial conditions. The figure demonstrates that the most and the least elongated initial voids tend to evolve in two 
opposite directions on $b/c$ axis: most elongated become less elongated in the course of evolution and least elongated 
become more elongated (see the black arrows indicating direction of the evolution). This effectively suppresses both tails of the shape distribution 
over time. However, loss in the tails of the distribution is simultaneously compensated by evolution of voids with initial shapes 
around the mean. These voids have equal chances of becoming more elongated or less elongated (see the black arrows). It appears 
that processes of suppression and compensation are so well tuned that the overall distribution of shapes does not evolve over time.

Considering an idealized model describing evolution of an isolated and aspherical void one can show that void should become 
gradually more spherical in the course of its evolution \citep{Icke1984}. As pointed out by \citet{Sha2006}, this reasoning is based 
on a major oversimplification. Voids do not evolve as isolated objects, but as a network of interconnected objects in which evolution of every individual void is constrained 
by evolution of its adjacent voids. The collective evolution of voids prevents them from gaining more spherical shapes. Our results 
reinforce this argument. As shown in Fig.~\ref{shape-z}, individual voids change their shapes with no global tendency towards 
sphericity. The mean axial ratio $a/c$ becomes actually slightly smaller indicating evolution towards more elongated shapes. 
As far as evolution of individual voids, our results show that a randomly selected void in initial conditions has equal chances 
of becoming more elongated or more spherical.

Our results demonstrate that non-linear evolution has a strong impact on shapes of individual voids. This should be 
regarded as a caveat for any attempt to model the distribution of void shapes using linear perturbation theory applied to 
the Gaussian random density field \citep[see e.g.][]{Lee2005}.

\subsection{Alignment}

\begin{figure}
\centering
\includegraphics[width=0.49\textwidth]{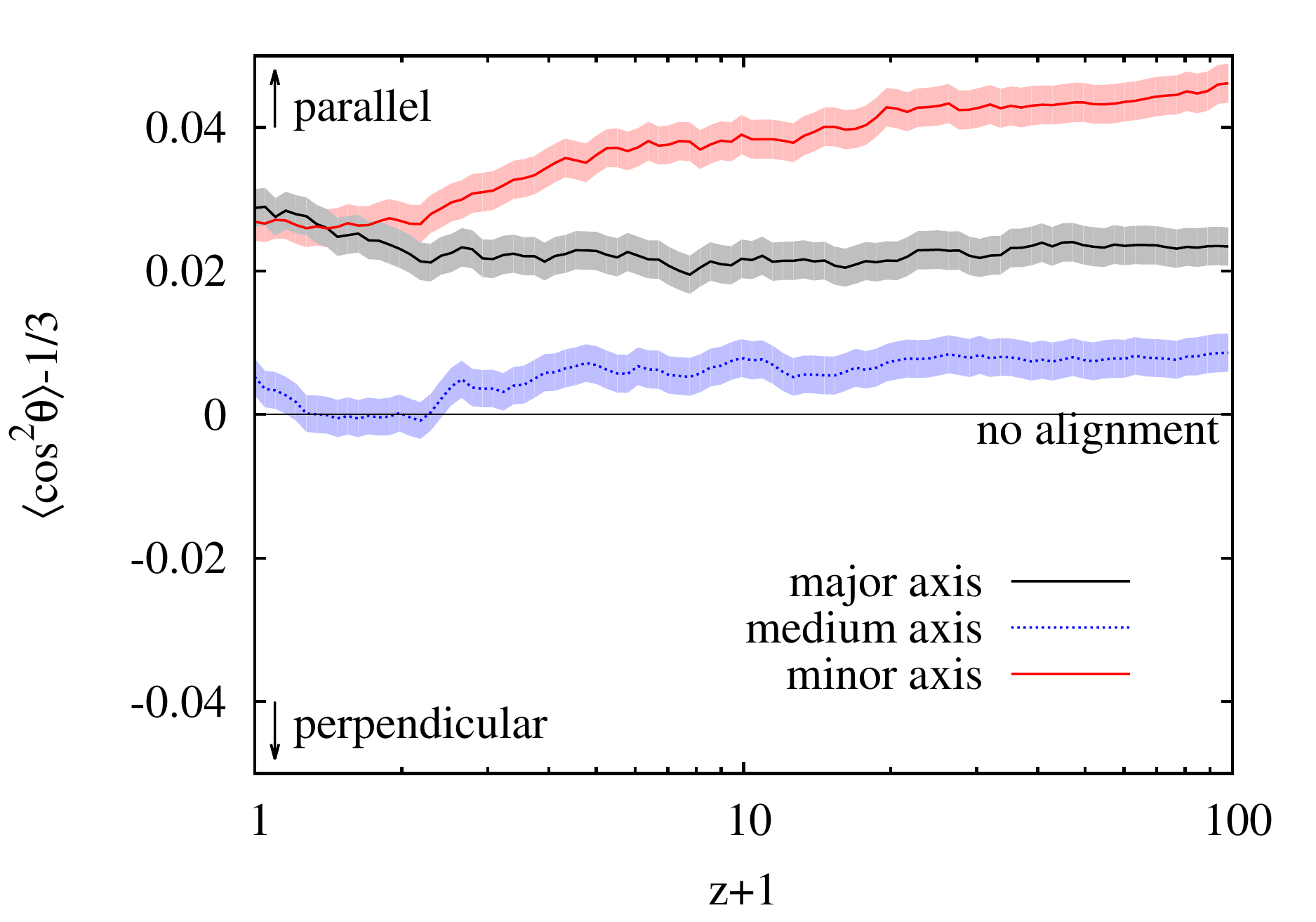}
\caption{Alignment of voids as a function of redshift. The alignment is measured by the angle 
$\theta$ between the corresponding principle axes of voids separated 
by no more than $15\hMpc$ (and excluding void-subvoid pairs). Positive 
or negative measure of the alignment shown on y-axis correspond 
to preferentially parallel or perpendicular orientations. The shaded bands 
show the $1\sigma$ errors from bootstrapping. Voids 
tend to be aligned in major and minor axes. This alignment is in large 
part arranged in the Gaussian primordial density field and only weakly 
modified in late-time evolution.
}
\label{alignment-z}
\end{figure}

Orientations of voids at redshift $z=0$ are strongly correlated at separations up to $20\hMpc$ \citep{Pla2008}. This 
kind of alignment is expected for all structures formed around peaks (or troughs for voids) of the Gaussian random density field 
\citep{Des2008}. Here we study to what degree the present alignment of voids is arranged in the primordial density field 
of initial conditions and to what degree it is modified in a phase of non-linear evolution of structure formation.

\begin{figure}
\centering
\includegraphics[width=0.49\textwidth]{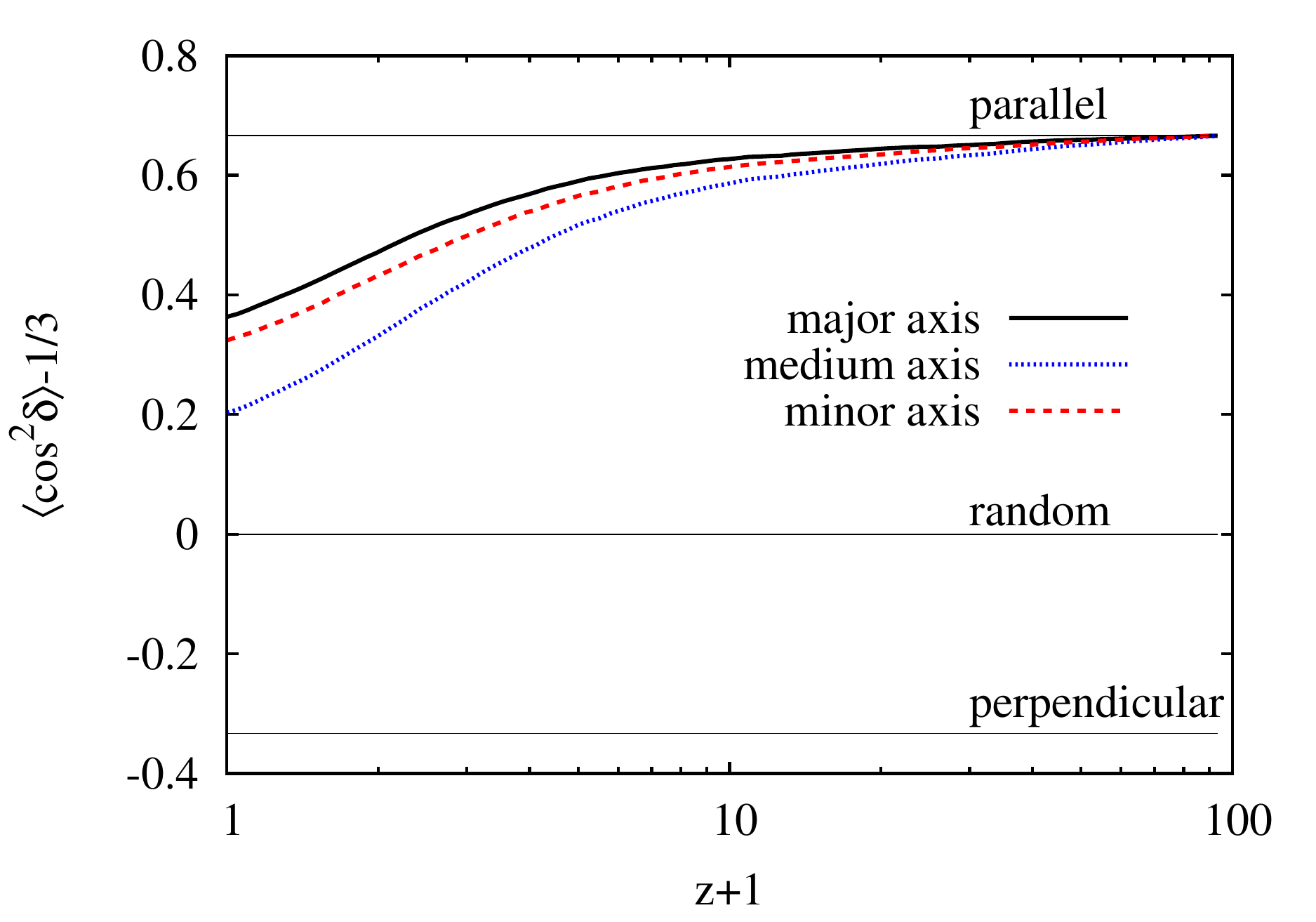}
\caption{Orientation of void principle axes as a function of redshift. 
The orientation is measured by the angle $\delta$ spanned 
by the principle axes from initial conditions to redshift $z$. Parallel configuration 
(no evolution of void orientations) corresponds to $2/3$, whereas random configuration 
(evolution entirely erasing initial orientations) to $0$. The orientations of void principle 
axes at late time retain a large part of the initial configuration.
}
\label{orient-z}
\end{figure}

We measure relative orientations of voids as the angle $\theta$ between the corresponding principle axes of 
voids in pairs. The principle axes are assumed to be eigenvectors of the shape tensor. We quantify degree of 
alignment by calculating $\langle \cos^{2}\theta\rangle-1/3$ which equals to $2/3$ for perfectly parallel orientations, 
$0$ for random orientations and $-1/3$ for perfectly perpendicular orientations. We consider void pairs with void 
separations less than $15\hMpc$ at which voids are maximally aligned at redshift $z=0$ \citep{Pla2008}. We 
use the same pairs of voids at all redshifts. In order to avoid potential biases due 
to hierarchical relations between voids, we discard all void-subvoid pairs.

Fig.~\ref{alignment-z} shows alignment of voids as a function of redshift. We find that voids at redshift $z=0$ exhibit strong and 
comparable alignment in major and minor axes, but nearly random orientations in medium axes, in full agreement with \citet{Pla2008}. 
Alignment of voids clearly undergoes some weak modifications in a phase of non-linear evolution. Compared to 
voids at $z=0$, voids at high redshifts (including initial conditions) are slightly more aligned in major axes than in medium axes. 
These differences, however, 
are quite small compared to the initial alignment at $z=100$. We therefore conclude that the current alignment of voids at $z=0$ is 
in large part determined by the Gaussian primordial density field.

Fig.~\ref{orient-z} reinforces our argument laid out above. It shows redshift evolution of void orientations measured by 
the angle $\delta$ spanned by the principle axes over time from initial conditions to redshift $z$. The evolution appears 
not to erase the initial orientations of void principle axes. The orientations of voids at late time retain a large part of the 
initial configuration acquired in the primordial density field. The correlation between the initial and final principle axes 
is far stronger than mutual alignments of void principle axes shown in Fig.~\ref{alignment-z}.
\subsection{Matter density distribution}

\begin{figure}
\centering
\includegraphics[width=0.49\textwidth]{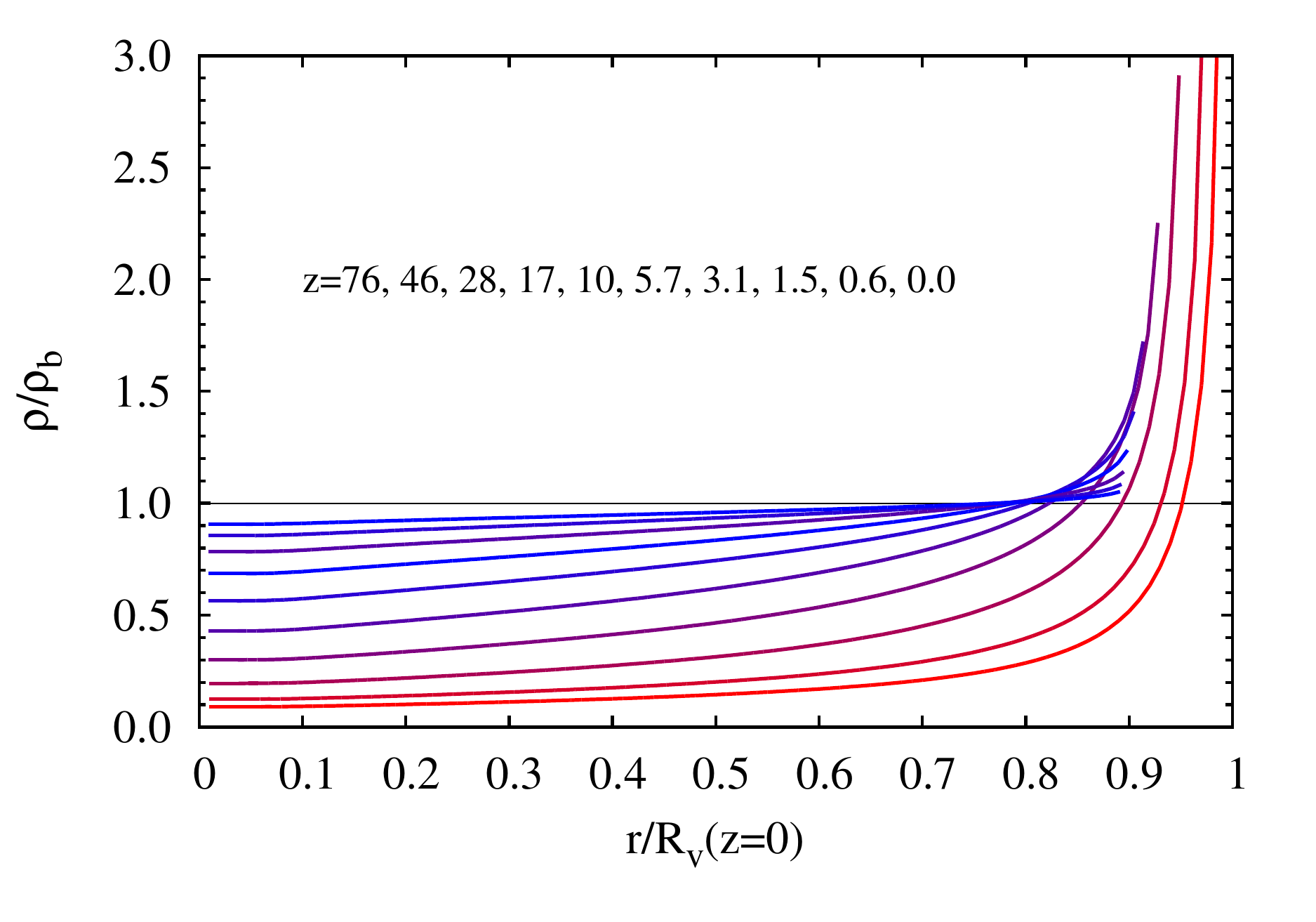}
\caption{Redshift evolution of the mean matter density distribution inside voids. The profiles relate densities 
of isodensity surfaces to effective radii $r$ of volumes enclosed by them. Using isodensity 
surfaces complies with the actual shapes of voids. The redshift sequence 
of the density distributions demonstrates formation of the final density profile 
with a characteristic bucket-like shape, i.e. nearly flat profile at $r\lesssim 0.9R_{\rm v}$ and 
a sharp transition to the walls at $r\approx R_{\rm v}$. The selected 
voids have effective radii $6\hMpc<R_{\rm v}<20\hMpc$ at $z=0$.
}
\label{rho-evolution}
\end{figure}

Most studies on the matter density distribution in voids assume spherical symmetry \citep[see e.g.][]{Ham2014, Sut2014dd, Sut2014b}. This approach is well justified 
when considering composite voids made by stacking individual objects together. On the other hand, averaging in spherical 
shells seems to be inappropriate for measuring matter density distribution in individual voids due to a high degree of 
asphericity. Capturing the true physical matter distribution requires using geometry which is consistent with the actual shapes of voids, 
e.g. aspherical surfaces whose points are at equal distances from the void boundary. As shown by \citet{Cau2015b} and \citet{Cau2015}, 
this approach changes dramatically the picture conveyed by the spherical density profiles. The new matter density profiles 
have far more extended flat parts in void cores and exhibit a sharp transition to high density walls on void boundaries. 
Interestingly, these features are fully consistent with simple analytic models describing formation of spherical and isolated voids 
\citep[see e.g.][]{She2004}.

Evolution of the matter density distribution in voids is driven in large part by matter evacuation in their cores and pilling 
up matter on their boundaries. Capturing how both processes develop relies on computing matter density distribution 
in a way which complies with actual shapes of voids. Here we adopt the following approach. We sort all pixels inside a given void 
according to their densities. Then we construct a density profile by finding effective radius $r$ of the volume given by 
all pixels with densities smaller than density $\rho(r)$. The resulting profile approximates closely a relation between 
densities of isodensity surfaces and effective radii of volumes enclosed by them. The choice of isodensity surfaces 
is a reasonable alternative to aspherical shells defined by equal distances from the void boundary \citep{Cau2015b}. 
As we shall see, both approaches give consistent results in terms of averaged density profiles.

Fig.~\ref{rho-evolution} shows the evolution of the mean density profile. The redshift sequence of the profiles demonstrates 
clearly two main processes shaping the final density distribution in $z=0$ voids: evacuation of matter in cores and formation 
of walls on the boundaries. The profiles of voids at highest redshifts evolve linearly with a symmetric growth 
of underdense and overdense parts. Intermediate redshifts show a transition to non-linear regime which breaks the symmetry 
between evolution of cores and walls. The final voids at redshift $z=0$ develop eventually the density profiles with a characteristic 
bucket-like shape: wide cores and a sharp transition to high-density walls. These profiles resemble remarkably well 
analytic models describing evolution of a spherical and isolated void \citep[see][]{She2004}. This resemblance is another justification 
that density profiles computed in a way which is consistent with the actual shapes of voids capture more physical properties 
than spherical density profiles.

\begin{figure}
\centering
\includegraphics[width=0.49\textwidth]{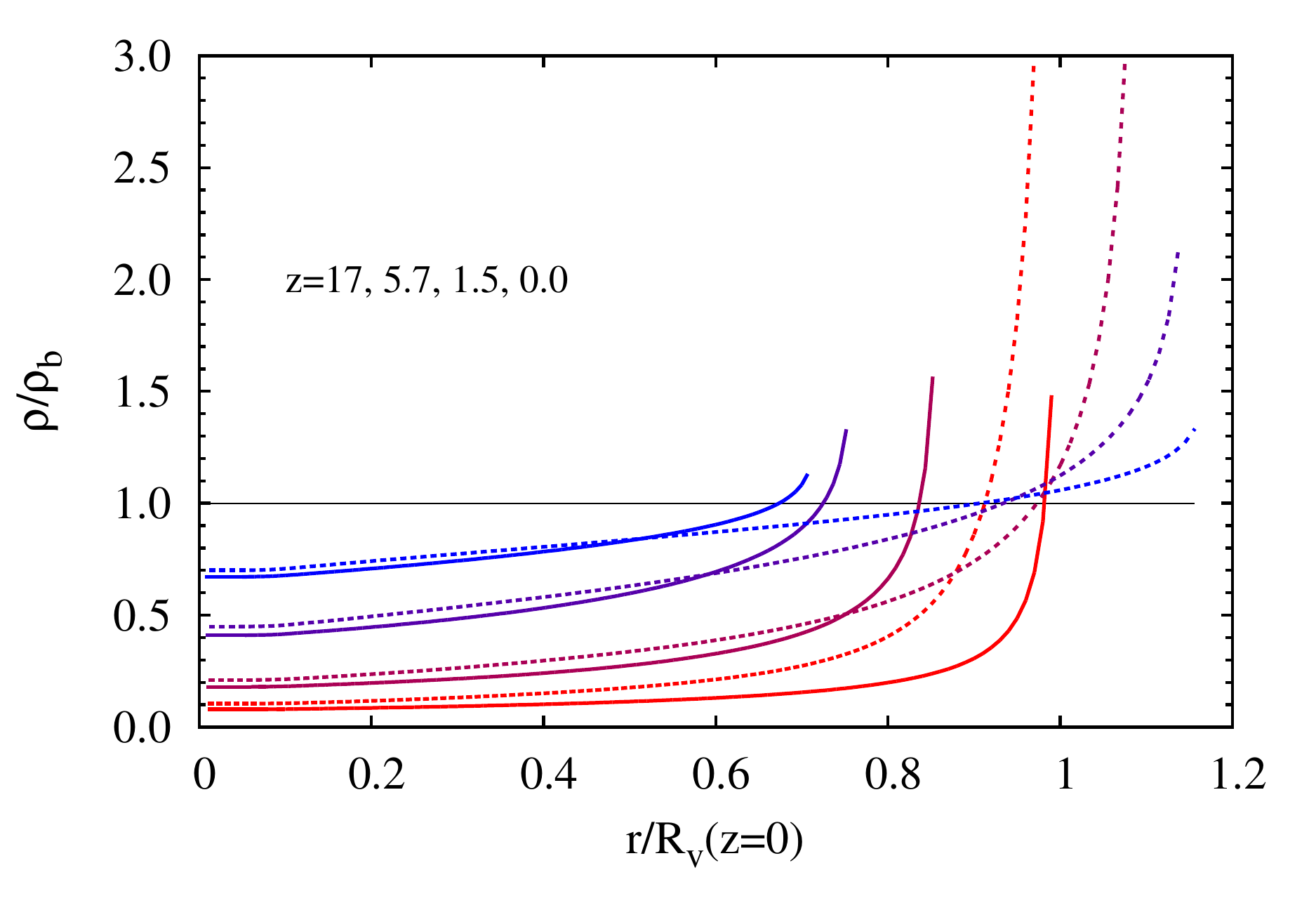}
\caption{Redshift evolution of the mean density distribution inside $10$ per cent 
of most expanding voids (solid lines) and $10$ per cent of most contracting voids (dashed lines). 
The profiles relate densities of isodensity surfaces to effective radii $r$ 
of volumes enclosed by them. Using isodensity surfaces complies with the actual shapes 
of voids and their progenitors. Despite 
distinct evolutions of the boundaries around the two types of voids, the cores undergo 
similar processes of matter evacuation. The expanding voids develop thinner walls, 
more extended cores part and a sharper transition between the core and the boundary walls. 
The selected voids have effective radii $6\hMpc<R_{\rm v}<20\hMpc$ 
at $z=0$.
}
\label{rho-ec-evolution}
\end{figure}

The density profiles at low redshifts demonstrate effect of void expansion already shown in Fig.~\ref{rad-evolution}. Since 
evolution of void sizes can substantially differ between individual objects, it is instructive to split the void sample according 
to their expansion histories. 
Fig.~\ref{rho-ec-evolution} shows evolution of the mean density profiles in two extreme groups of voids: $10$ per cent of 
most expanding and $10$ per cent of most contracting voids (see the tails of the distribution in Fig.~\ref{rad-hierarchy}). 
The figure clearly demonstrates that expansion or contraction of void boundaries is independent of how the core evolves. 
It is directly related to how massive walls are developed around voids: voids with 
massive walls can undergo contraction, whereas voids with weak walls tend to expand. It is therefore not surprising that this 
relation reflects also levels of the void hierarchy. Subvoids develop weaker walls and their 
boundaries are dragged by super-Hubble flow of their host voids. On the other hand, voids at the top level of the hierarchy 
pile up massive walls which can likely start to collapse and shrink void interiors.

The two groups of most expanding and most contracting voids develop slightly different density profiles. The expanding 
voids (bottom levels in the void hierarchy) are characterized by larger cores, weaker walls and a sharper transition between the 
core and the boundaries. On the other hand, the contracting voids (top levels in the void hierarchy) develop more massive 
and thicker walls with a smoother transition between the core and the boundaries.

\section{Summary and conclusions}

We have analysed performance of two different methods of tracking voids in cosmological simulations. The first 
approach is based on a direct application of a halo merger tree generator to void catalogues. Void tracking is performed 
by means of matching voids between successive snapshots of the simulation. The second approach has been developed 
for the purpose of this work and it finds voids in every snapshot of the simulation by tracking watershed basins of 
well formed voids. While the former method involves generating void catalogue at all snapshot, the latter builds the void catalogue only once.

We have shown that the method based on matching voids between successive snapshots gives rise to spurious instantaneous 
reconfigurations of voids. These effects are not physical and they are related to persistent reordering of minima 
and saddle points in the simulated density field, in large part due to unresolved small-scale modes. We have demonstrated 
that the new method introduced in this work eliminates these problems and yields a continuous mapping between boundaries 
of voids at all redshift, from initial conditions to the present time.

We have applied the developed method of void tracking to cosmological simulations of a standard $\Lambda$CDM model. 
This has allowed us to study for the first time the redshift evolution of various properties of individual voids. We have restricted 
our studies to voids with effective radii between $6\hMpc$ and $20\hMpc$ at redshift $z=0$. Our results can be summarized as follows.

\begin{description}
\item[i.] Initial void boundaries undergo complicated distortions caused by tidal forces acting in a late-time phase of the evolution. 
The spatial scale of these deformations is comparable to void sizes.

\item[ii.] Evolution of void volumes is far less prominent than the boundary surfaces. On average voids increase their volumes by only 
$15$ per cent. Individual voids can either expand or contract (collapse) depending on their location in the void hierarchy: expanding 
voids lie at the bottom of the hierarchy (subvoids), whereas collapsing voids at the top of the hierarchy. This connection 
is an additional factor modulating a well-known relationship between expansion rate of voids and their sizes (large voids tend 
to expand and small voids tend to contract or collapse).

\item[iii.] Individual voids change their initial shapes acquired in the primordial density field. Evolution of void shapes proceeds in a way 
which barely modifies the overall distribution of axis ratios. This happens due to the fact that most elongated voids tend to become more 
spherical and the most spherical ones develop more elongated shapes. There is a weak trend of the major-to-minor axis ratio to 
decrease in time (developing gradually more elongated shapes of voids).

\item[iv.] Evolution of voids does not erase their initial orientations. The orientations of void principle axes at late time are highly 
correlated with the initial ones.

\item[v.] Major and minor axes of voids tend to be aligned at all redshifts. Lack of significant evolution suggests that void alignment origins 
in large part from the Gaussian primordial density field.

\item[vi.] The new method of tracking voids in cosmological simulations allows us to capture redshift evolution of the matter density 
distribution in every individual void. Evolution of the matter density profiles in voids, computed on isodensity surfaces shows the 
formation of theoretically predicted profiles with a characteristic bucket-like shape indicating a vast core of nearly constant density 
and a sharp transition to high-density walls.

\item[vii.] The matter evacuation in void cores is decoupled from the evolution of void boundaries. Evolution of void cores 
proceeds in the same way both in expanding and contracting voids. Unlike voids with boundaries dragged by super-Hubble flow, 
voids undergoing contraction tend to pile up thicker walls around them and a slightly smoother transition between the core and 
void boundary.

\end{description}

Despite clearly apparent evolution of every individual void, we conclude that the final voids do not substantially differ from their initial 
state in many respects. Apart from obvious difference in terms of the core density, voids retain a large part of properties acquired 
in the primordial density field such as shapes and orientations. It is even more surprising to notice that evolutionary paths of individual 
voids conspire to conceal any signature of evolution in overall statistics of voids. In particular, the evolution of axial ratios quantifying 
void shapes barely modifies the overall distribution over time. This reinforces the argument that tracing the evolution of individual 
voids in cosmological simulations unveils far more detailed aspects of void evolution than what one can learn from measuring overall 
distributions of various void properties in a range of redshifts.

\section*{Acknowledgements}
RW would like to thank Rien van de Weygaert for useful discussions, Marius Cautun for critical reading of the manuscript 
and Ralf Kaehler for his help on 3D data visualization. The authors gratefully acknowledge the anonymous referee for constructive comments. 
RW acknowledges support through the Porat Postdoctoral Fellowship. The Dark Cosmology 
Centre is funded by the Danish National Research Foundation. 

\bibliography{master}

\begin{thebibliography}{51}
\expandafter\ifx\csname natexlab\endcsname\relax\def\natexlab#1{#1}\fi

\bibitem[{{Abel} {et~al}\mbox{.}(2012){Abel}, {Hahn}, \& {Kaehler}}]{Abel2012}
{Abel} T., {Hahn} O., {Kaehler} R., 2012, \mnras, 427, 61

\bibitem[{{Angulo} {et~al}\mbox{.}(2014){Angulo}, {Chen}, {Hilbert}, \&
  {Abel}}]{Ang2014}
{Angulo} R.~E., {Chen} R., {Hilbert} S., {Abel} T., 2014, \mnras, 444, 2925

\bibitem[{{Aragon-Calvo} \& {Szalay}(2013)}]{Ara2013}
{Aragon-Calvo} M.~A., {Szalay} A.~S., 2013, \mnras, 428, 3409

\bibitem[{{Aragon-Calvo} {et~al}\mbox{.}(2010){Aragon-Calvo}, {van de
  Weygaert}, {Araya-Melo}, {Platen}, \& {Szalay}}]{Ara2010}
{Aragon-Calvo} M.~A., {van de Weygaert} R., {Araya-Melo} P.~A., {Platen} E.,
  {Szalay} A.~S., 2010, \mnras, 404, L89

\bibitem[{{Bond} {et~al}\mbox{.}(1996){Bond}, {Kofman}, \&
  {Pogosyan}}]{Bon1996}
{Bond} J.~R., {Kofman} L., {Pogosyan} D., 1996, \nat, 380, 603

\bibitem[{{Bos} {et~al}\mbox{.}(2012){Bos}, {van de Weygaert}, {Dolag}, \&
  {Pettorino}}]{Bos2012}
{Bos} E.~G.~P., {van de Weygaert} R., {Dolag} K., {Pettorino} V., 2012, \mnras,
  426, 440

\bibitem[{{Cai} {et~al}\mbox{.}(2015){Cai}, {Padilla}, \& {Li}}]{Cai2015}
{Cai} Y.-C., {Padilla} N., {Li} B., 2015, \mnras, 451, 1036

\bibitem[{{Cautun} {et~al}\mbox{.}(2016){Cautun}, {Cai}, \& {Frenk}}]{Cau2015b}
{Cautun} M., {Cai} Y.-C., {Frenk} C.~S., 2016, \mnras, 457, 2540

\bibitem[{{Cautun} {et~al}\mbox{.}(2013){Cautun}, {van de Weygaert}, \&
  {Jones}}]{Cau2013}
{Cautun} M., {van de Weygaert} R., {Jones} B.~J.~T., 2013, \mnras, 429, 1286

\bibitem[{{Cautun} {et~al}\mbox{.}(2014){Cautun}, {van de Weygaert}, {Jones},
  \& {Frenk}}]{Cau2014}
{Cautun} M., {van de Weygaert} R., {Jones} B.~J.~T., {Frenk} C.~S., 2014,
  \mnras, 441, 2923

\bibitem[{{Cautun} {et~al}\mbox{.}(2015){Cautun}, {van de Weygaert}, {Jones},
  \& {Frenk}}]{Cau2015}
{Cautun} M., {van de Weygaert} R., {Jones} B.~J.~T., {Frenk} C.~S., 2015, ArXiv
  e-prints 1501.01306

\bibitem[{{Colberg} {et~al}\mbox{.}(2008){Colberg}, {Pearce}, {Foster},
  {Platen}, {Brunino}, {Neyrinck}, {Basilakos}, {Fairall}, {Feldman},
  {Gottl{\"o}ber}, {Hahn}, {Hoyle}, {M{\"u}ller}, {Nelson}, {Plionis},
  {Porciani}, {Shandarin}, {Vogeley}, \& {van de Weygaert}}]{Col2008}
{Colberg} J.~M. {et~al.}, 2008, \mnras, 387, 933

\bibitem[{{Desjacques} \& {Smith}(2008)}]{Des2008}
{Desjacques} V., {Smith} R.~E., 2008, \prd, 78, 023527

\bibitem[{{Eisenstein} \& {Hu}(1998)}]{Eis1998}
{Eisenstein} D.~J., {Hu} W., 1998, \apj, 496, 605

\bibitem[{{Falck} \& {Neyrinck}(2015)}]{Fal2015}
{Falck} B., {Neyrinck} M.~C., 2015, \mnras, 450, 3239

\bibitem[{{Gottl{\"o}ber} {et~al}\mbox{.}(2003){Gottl{\"o}ber}, {{\L}okas},
  {Klypin}, \& {Hoffman}}]{Got2003}
{Gottl{\"o}ber} S., {{\L}okas} E.~L., {Klypin} A., {Hoffman} Y., 2003, \mnras,
  344, 715

\bibitem[{{Hahn} \& {Abel}(2011)}]{Hahn2011}
{Hahn} O., {Abel} T., 2011, \mnras, 415, 2101

\bibitem[{{Hahn} {et~al}\mbox{.}(2013){Hahn}, {Abel}, \& {Kaehler}}]{Hah2013}
{Hahn} O., {Abel} T., {Kaehler} R., 2013, \mnras, 434, 1171

\bibitem[{{Hahn} \& {Angulo}(2016)}]{Hah2016}
{Hahn} O., {Angulo} R.~E., 2016, \mnras, 455, 1115

\bibitem[{{Hahn} {et~al}\mbox{.}(2015){Hahn}, {Angulo}, \& {Abel}}]{Hah2015}
{Hahn} O., {Angulo} R.~E., {Abel} T., 2015, \mnras, 454, 3920

\bibitem[{{Hahn} {et~al}\mbox{.}(2007){Hahn}, {Porciani}, {Carollo}, \&
  {Dekel}}]{Hah2007}
{Hahn} O., {Porciani} C., {Carollo} C.~M., {Dekel} A., 2007, \mnras, 375, 489

\bibitem[{{Hamaus} {et~al}\mbox{.}(2014){Hamaus}, {Sutter}, \&
  {Wandelt}}]{Ham2014}
{Hamaus} N., {Sutter} P.~M., {Wandelt} B.~D., 2014, Physical Review Letters,
  112, 251302

\bibitem[{{Hellwing} \& {Juszkiewicz}(2009)}]{Hel2009}
{Hellwing} W.~A., {Juszkiewicz} R., 2009, \prd, 80, 083522

\bibitem[{Hockney \& Eastwood(1988)}]{Hoc1988}
Hockney R.~W., Eastwood J.~W., 1988, Computer Simulation Using Particles.
  Taylor \& Francis, Inc., Bristol, PA, USA

\bibitem[{{Hoffman} {et~al}\mbox{.}(2012){Hoffman}, {Metuki}, {Yepes},
  {Gottl{\"o}ber}, {Forero-Romero}, {Libeskind}, \& {Knebe}}]{Hof2012}
{Hoffman} Y., {Metuki} O., {Yepes} G., {Gottl{\"o}ber} S., {Forero-Romero}
  J.~E., {Libeskind} N.~I., {Knebe} A., 2012, \mnras, 425, 2049

\bibitem[{{Icke}(1984)}]{Icke1984}
{Icke} V., 1984, \mnras, 206, 1P

\bibitem[{Kaehler {et~al}\mbox{.}(2012)Kaehler, Hahn, \& Abel}]{Kae2012}
Kaehler R., Hahn O., Abel T., 2012, IEEE Transactions on Visualization and
  Computer Graphics, 18, 2078

\bibitem[{{Lambas} {et~al}\mbox{.}(2016){Lambas}, {Lares}, {Ceccarelli},
  {Ruiz}, {Paz}, {Maldonado}, \& {Luparello}}]{Lam2016}
{Lambas} D.~G., {Lares} M., {Ceccarelli} L., {Ruiz} A.~N., {Paz} D.~J.,
  {Maldonado} V.~E., {Luparello} H.~E., 2016, \mnras, 455, L99

\bibitem[{{Lavaux} \& {Wandelt}(2012)}]{Lav2012}
{Lavaux} G., {Wandelt} B.~D., 2012, \apj, 754, 109

\bibitem[{{Lee} {et~al}\mbox{.}(2005){Lee}, {Jing}, \& {Suto}}]{Lee2005}
{Lee} J., {Jing} Y.~P., {Suto} Y., 2005, \apj, 632, 706

\bibitem[{{Li} {et~al}\mbox{.}(2012){Li}, {Zhao}, \& {Koyama}}]{Li2012}
{Li} B., {Zhao} G.-B., {Koyama} K., 2012, \mnras, 421, 3481

\bibitem[{{Neyrinck}(2008)}]{Ney2008}
{Neyrinck} M.~C., 2008, \mnras, 386, 2101

\bibitem[{{Paz} {et~al}\mbox{.}(2013){Paz}, {Lares}, {Ceccarelli}, {Padilla},
  \& {Lambas}}]{Paz2013}
{Paz} D., {Lares} M., {Ceccarelli} L., {Padilla} N., {Lambas} D.~G., 2013,
  \mnras, 436, 3480

\bibitem[{{Planck Collaboration} {et~al}\mbox{.}(2014){Planck Collaboration},
  {Ade}, {Aghanim}, {Armitage-Caplan}, {Arnaud}, {Ashdown}, {Atrio-Barandela},
  {Aumont}, {Baccigalupi}, {Banday}, \& et~al.}]{Planck2014a}
{Planck Collaboration} {et~al.}, 2014, \aap, 571, A16

\bibitem[{{Platen} {et~al}\mbox{.}(2007){Platen}, {van de Weygaert}, \&
  {Jones}}]{Pla2007}
{Platen} E., {van de Weygaert} R., {Jones} B.~J.~T., 2007, \mnras, 380, 551

\bibitem[{{Platen} {et~al}\mbox{.}(2008){Platen}, {van de Weygaert}, \&
  {Jones}}]{Pla2008}
{Platen} E., {van de Weygaert} R., {Jones} B.~J.~T., 2008, \mnras, 387, 128

\bibitem[{Powell \& Abel(2015)}]{Pow2015}
Powell D., Abel T., 2015, Journal of Computational Physics, 297, 340

\bibitem[{{Sahni} {et~al}\mbox{.}(1994){Sahni}, {Sathyaprakah}, \&
  {Shandarin}}]{Sah1994}
{Sahni} V., {Sathyaprakah} B.~S., {Shandarin} S.~F., 1994, \apj, 431, 20

\bibitem[{{Shandarin} {et~al}\mbox{.}(2006){Shandarin}, {Feldman}, {Heitmann},
  \& {Habib}}]{Sha2006}
{Shandarin} S., {Feldman} H.~A., {Heitmann} K., {Habib} S., 2006, \mnras, 367,
  1629

\bibitem[{{Shandarin} {et~al}\mbox{.}(2012){Shandarin}, {Habib}, \&
  {Heitmann}}]{Sha2012}
{Shandarin} S., {Habib} S., {Heitmann} K., 2012, \prd, 85, 083005

\bibitem[{{Sheth} \& {van de Weygaert}(2004)}]{She2004}
{Sheth} R.~K., {van de Weygaert} R., 2004, \mnras, 350, 517

\bibitem[{{Sousbie} \& {Colombi}(2015)}]{Sou2015}
{Sousbie} T., {Colombi} S., 2015, ArXiv e-prints 1509.07720

\bibitem[{{Springel}(2005)}]{Spr2005}
{Springel} V., 2005, \mnras, 364, 1105

\bibitem[{{Sutter} {et~al}\mbox{.}(2014{\natexlab{a}}){Sutter}, {Elahi},
  {Falck}, {Onions}, {Hamaus}, {Knebe}, {Srisawat}, \& {Schneider}}]{Sut2014c}
{Sutter} P.~M., {Elahi} P., {Falck} B., {Onions} J., {Hamaus} N., {Knebe} A.,
  {Srisawat} C., {Schneider} A., 2014{\natexlab{a}}, \mnras, 445, 1235

\bibitem[{{Sutter} {et~al}\mbox{.}(2015){Sutter}, {Lavaux}, {Hamaus}, {Pisani},
  {Wandelt}, {Warren}, {Villaescusa-Navarro}, {Zivick}, {Mao}, \&
  {Thompson}}]{Sut2015}
{Sutter} P.~M. {et~al.}, 2015, Astronomy and Computing, 9, 1

\bibitem[{{Sutter} {et~al}\mbox{.}(2014{\natexlab{b}}){Sutter}, {Lavaux},
  {Hamaus}, {Wandelt}, {Weinberg}, \& {Warren}}]{Sut2014dd}
{Sutter} P.~M., {Lavaux} G., {Hamaus} N., {Wandelt} B.~D., {Weinberg} D.~H.,
  {Warren} M.~S., 2014{\natexlab{b}}, \mnras, 442, 462

\bibitem[{{Sutter} {et~al}\mbox{.}(2014{\natexlab{c}}){Sutter}, {Lavaux},
  {Wandelt}, {Weinberg}, \& {Warren}}]{Sut2014b}
{Sutter} P.~M., {Lavaux} G., {Wandelt} B.~D., {Weinberg} D.~H., {Warren} M.~S.,
  2014{\natexlab{c}}, \mnras, 438, 3177

\bibitem[{{van de Weygaert} \& {Platen}(2011)}]{Wey2011}
{van de Weygaert} R., {Platen} E., 2011, International Journal of Modern
  Physics Conference Series, 1, 41

\bibitem[{{van de Weygaert} \& {van Kampen}(1993)}]{Wey1993}
{van de Weygaert} R., {van Kampen} E., 1993, \mnras, 263, 481

\bibitem[{{Way} {et~al}\mbox{.}(2015){Way}, {Gazis}, \& {Scargle}}]{Way2015}
{Way} M.~J., {Gazis} P.~R., {Scargle} J.~D., 2015, \apj, 799, 95

\bibitem[{{Yang} {et~al}\mbox{.}(2015){Yang}, {Neyrinck}, {Arag{\'o}n-Calvo},
  {Falck}, \& {Silk}}]{Yan2015}
{Yang} L.~F., {Neyrinck} M.~C., {Arag{\'o}n-Calvo} M.~A., {Falck} B., {Silk}
  J., 2015, \mnras, 451, 3606

\end{thebibliography}

\appendix

\end{document}